\documentclass[aps,pre,twocolumn,groupedaddress]{revtex4-2}
\usepackage{latexsym}
\usepackage{amsmath}
\usepackage{amsfonts}
\usepackage{amssymb}
\usepackage{amscd}
\usepackage{bbm}
\usepackage{bbold}
\usepackage{tikz}
\usepackage{tikz-cd}
\usetikzlibrary{shapes}
\usepackage{hyperref}
\usepackage[export]{adjustbox}
\usepackage[caption=false]{subfig}
\usepackage{algorithm}
\usepackage[noend]{algpseudocode}

\newcommand{\Z}{\mathcal{Z}}
\newcommand{\ket}[1]{|#1\rangle}
\newcommand{\bra}[1]{\langle#1|}
\newcommand{\bracket}[2]{\langle#1|#2\rangle}
\newcommand{\vev}[1]{\langle#1\rangle}

\newcommand{\SI}{\textsc{s}\textsc{i}}
\renewcommand{\S}{\textsc{s}}
\newcommand{\I}{\textsc{i}}

\newcommand{\qSI}{\mathfrak{q}_{\S \to \I}}

\newcommand{\Q}{\mathbb{Q}}

\begin{document}

\title{Logistic growth on networks: exact solutions for the SI model}

\author{Wout Merbis}
\email[]{w.merbis@uva.nl}
\affiliation{Dutch Institute for Emergent Phenomena (DIEP), Institute for Theoretical Physics, University of Amsterdam, 1090 GL Amsterdam, The Netherlands}

\author{Ivano Lodato}
\email[]{ivano.lodato@gmail.com}
\affiliation{Allos Limited, Hong Kong}

\date{\today}

\begin{abstract}
The susceptible-infected (SI) model is the most basic of all compartmental models used to describe the spreading of information through a population. Despite its apparent simplicity, the analytic solution of this model on networks is still lacking.
We address this problem here, using a novel formulation inspired by the mathematical treatment of many-body quantum systems. 
This allows us to organize the time-dependent expectation values for the state of individual nodes in terms of contributions from subgraphs of the network. We compute these contributions systematically and find a set of symmetry relations among subgraphs of differing topologies.
We use our novel approach to compute the spreading of information on three different sample networks. The exact solution, which matches with Monte-Carlo simulations, visibly departs from the mean-field results. 
\end{abstract}

\maketitle

\section{Introduction}
The dynamical processes of information spreading through a population via individual interaction is ubiquitous in Nature and society \cite{Anderson1992,strogatz2001exploring, albert2002statistical, boccaletti2006complex, barrat2008dynamical, castellano2009statistical, newman2018networks}, and their importance is demonstrated by the recent COVID-19 pandemic \cite{carletti2020covid,nielsen2021covid,moein2021inefficiency}.
Typically, there are two approaches to formulate spreading processes: a collective approach based on deterministic \emph{compartmental models} governing the state of the population as a whole \cite{kermack1927,Anderson1992,Brauer2008,keeling2011modeling} and a more detailed approach aimed at describing the stochastic individual interactions \cite{pastor2001epidemic,Newman2002,PastorSatorras2015}.  
Various approximation and numerical schemes have been used to solve compartmental models of epidemic spreading, and in certain cases, exact results are known \cite{khan2009explicit,harko2014exact,bohner2019exact,kozyreff2021near}. These models are effective and often give a good estimate of the temporal evolution of the system. However, they are insensitive to the connectivity properties of the individuals within the population. To take this into account, stochastic models of epidemic spreading on networks have been proposed and studied extensively over the past decades \cite{pastor2001epidemic,may2001infection,pastor2002epidemic,eguiluz2002epidemic,boguna2003absence,van2008virus,castellano2010thresholds,van2011n,PastorSatorras2015,cai2016solving,Wang_2017,harding2018thermodynamic,matamalas2018effective,wang2019coevolution}. Nonetheless, the presence of dynamical correlations and the non-linear nature of spreading processes make mean-field approximations and numerical simulations a necessity, while exact solutions in network epidemiology remain a rare treat \cite{simon2011exact,kiss2015exact,cator2013susceptible,prasse2020time}.

In this paper we present a general framework to describe the spreading process on a network, which allows for \emph{exact analytic solutions}. We focus on the most basics and yet relevant \cite{demongeot2020si,moore2020predicting} of all spreading processes, the $\SI$ system. The nodes of the network are either susceptible ($\S$) or infected ($\I$) and, by a Poisson process, infected nodes can infect their susceptible neighbors. While the compartmental model for this system is solved by the Verhulst logistic function \cite{verhulst1838}, to the best of our knowledge, exact solutions to the stochastic network model are still unavailable. Here we intend to fill this gap, while simultaneously introducing a general procedure applicable to other, more complex, spreading processes.

Our exact treatment of the $\SI$ system relies on the formulation of \cite{merbis2021exact}, inspired by many-body quantum systems. Time-dependent expectation values of target nodes (sinks), given an initial configuration of infected nodes (sources), can be obtained as a diagrammatic expansion over subgraphs of the network.
Each subgraph represents a contribution to the spreading process from the sources to the sinks.
We present a systematic way to compute these contributions as functions of time and furthermore, we find a set of symmetry relations among subgraphs of different topology, which simplifies the computational task. 
We illustrate our methods by analytically solving the spreading dynamics on three sample networks. The result differs visibly from the individual-based mean-field approximation, but agrees with Monte-Carlo simulations.

\section{The formalism} 
Consider an unweighted and undirected network of $N$ nodes, each of which can be in one of two states: $\S$ or $\I$. 
These states can be used as an orthonormal basis for a two-dimensional (2D) vector space,  such that \footnote{We use a bra-ket notation for column vectors $\bra{a}$ and row vectors $\ket{b}$, such that their contraction is denoted as $\bracket{a}{b}$. }:
\begin{equation}\label{statevec}
    \ket{\S} = \begin{pmatrix}
           1\\
           0
         \end{pmatrix} \,, \qquad \ket{\I} = \begin{pmatrix}
           0\\
           1
         \end{pmatrix} \,.
\end{equation}  
The network as a whole can be in $2^N$ possible configurations $\mathcal{C} \in \{\S,\I\}^N$, each of which has an associated vector $\ket{\mathcal{C}}$ constructed as the tensor product of $N$ individual state vectors \eqref{statevec} 
\begin{equation}
\ket{\mathcal{C}} = \ket{1}\otimes \ket{2} \otimes \dots \otimes \ket{N}
\end{equation}
where each state $\ket{i}=\{\ket{\S},\ket{\I}$\}. 
The probabilities $p(\mathcal{C},t)$ for the network to be in configuration $\mathcal{C}$ at time $t$ are the components of a $2^N$-dimensional probability vector $\ket{\rho(t)}$
\begin{equation}
\ket{\rho(t)} = \sum_\mathcal{C \in \{ \S,\I\}^N } p(\mathcal{C},t) \ket{\mathcal{C}}
\end{equation} 
This vector has unit $L^1$-norm by normalization of probability, which we will denote as $\bracket{1}{\rho(t)} = 1$, where $\bra{1} = \bigotimes_{i=1}^N (1 \; 1)$ denotes the flat state. The $\SI$ spreading is a Markovian process and hence the probability vector evolves in time by a master equation: 
\begin{equation}\label{mastereqn}
    \frac{d}{dt} \ket{\rho(t)} = \Q(A) \ket{\rho(t)}\,.
\end{equation}
The infinitesimal generator $\Q(A)$ is a $2^N$-dimensional matrix containing the possible transitions induced between neighboring states in a network with adjacency matrix $A$. It can be constructed from tensor products of 2D matrices, which act as linear operators on the individual state vectors. In the case of the $\SI$-model, it is given by \cite{merbis2021exact}:
\begin{equation}\label{QSI}
    \Q(A) = \sum_{i,j=1}^N A^{ij} P_{\I}^i \cdot \qSI^j \,.
\end{equation}
The $2^N$-dimensional matrices $P_\I^i$ and $\qSI^j$ are constructed by taking the tensor product of $N-1$ 2D identity matrices with the projection operator $P_\I$ 
inserted at site $i$ and the infinitesimal stochastic matrix $\qSI$ 
inserted at site $j$, respectively, and:
\begin{equation}\label{operatordef}
    P_\I =  \begin{pmatrix}
               0 & 0 \\
               0 & 1
    \end{pmatrix}\,, \qquad
    \qSI =  \begin{pmatrix}
               -1 & 0 \\
               1 & 0
    \end{pmatrix}\,.
\end{equation}
Each term in $\Q(A)$ performs the following local operations on the individual nodes $i,j$ connected by an edge: $P_\I$ checks whether node $i$ is infected, by projecting it to the infected basis vector $\ket{\I}$. The matrix $\qSI$ simultaneously acts on node $j$ and maps the state $\ket{\S}$ to the infinitesimal stochastic state $\ket{\I}-\ket{\S}$. This state with vanishing $L^1$ norm subtracts probability for the node $j$ to remain in the susceptible state and adds probability of it being infected.

Our goal is to compute the probability for a certain sink node $i$ to be infected at time $t$, given a fixed configuration of initially infected source nodes at time $t=0$. 
\begin{equation}\label{expvalue}
    \vev{\I^i (t)} = \bra{1} P_\I^i \ket{\rho(t)} \,.
\end{equation}
We will show in section \ref{sec:diagrams} below how this quantity can be expressed as a sum over subgraph diagrams with specific node configurations. Terms where any node other than $i$ are in the infinitesimal stochastic state $\ket{\I}-\ket{\S}$ will vanish due to the contraction with the flat state $\bra{1}$. Therefore, the function $\vev{\I^i (t)} $ will only receive contributions from terms where no node other than $i$ is in an infinitesimal stochastic state. We can hence identify the sinks of the spreading process with nodes in an infinitesimal stochastic state. Before proceeding to the diagrammatic expansion of \eqref{expvalue}, we will discuss how to this formula is derived from a dynamical partition function.

\section{Dynamical Partition function}
In this section, we will give an expression for the dynamical partition function \cite{gaspard2005chaos,lecomte2007thermodynamic} in the ensemble average over all possible trajectories of the spreading process. We will use this to derive equation \eqref{expvalue}, as well as  expressions for higher-order moments. 
The partition function is defined as the moment generating function for the observables $I^i(t)$, which tracks whether node $i$ is infected at time $t$. 
\begin{equation}\label{ZI}
    \Z_I(\{s_i\},t) = \left\langle \exp\left( \sum_{i=1}^N s_i I^i(t) \right) \right\rangle\,.
\end{equation}
Where here $I^i(t) = 1$ if node $i$ is infected at time $t$ and $I^i(t) = 0$ otherwise. The dynamical partition function depends on a set of $N$ (dual) variables $\{s_i\}$, which serve as chemical potentials for the observables $I^i(t)$. The expectation value $\vev{\I^{i_1}(t)\cdots \I^{i_n}(t)}$ for the joint probability of nodes $i_1, \ldots, i_n$ to be infected at time $t$ is obtained from \eqref{ZI} by derivation with respect to $s_{i_1}, \dots, s_{i_n}$, followed by setting all $\{s_i\} =0$:
\begin{equation}\label{expfromZ}
    \vev{\I^{i_1}(t)\cdots \I^{i_n}(t)} =  \partial_{s_{i_1}} \cdots \partial_{s_{i_n}} \Z_I(\{s_i\},t) |_{\{s_i\} = 0}\,.
\end{equation}
The dynamical partition function \eqref{ZI} can be computed by summing $e^{\sum_i s_i I^i(\mathcal{C})} p(\mathcal{C},t)$ over all microscopic configurations $\mathcal{C} \in \{\S,\I\}^N$, where $I^i(\mathcal{C})$ gives 1 only if node $i$ is infected in the configuration $\mathcal{C}$. In the vector notation $\ket{\rho(t)} = \sum_\mathcal{C} p(\mathcal{C},t) \ket{\mathcal{C}}$, we can express $e^{\sum_i s_i I^i(\mathcal{C})}$ as a linear operator $T(\{s_i\})$ which places a factor $e^{s_i}$ for each infected node $i$ in the configuration $\mathcal{C}$. This operator is a $2^N$-dimensional diagonal matrix with the following product form: 
\begin{equation}\label{Tproductform}
    T(\{s_i\}) = \exp\left( \sum_{i=1}^N s_i P_{\I}^i \right) = \bigotimes_{i=1}^N \left( \ket{\S}\bra{\S} + e^{s_i} \ket{\I}\bra{\I} \right)\,.
\end{equation}
Now \eqref{ZI} can be written as the $L^1$-norm of the contraction of $T(\{s_i\})$ with the probability vector $\ket{\rho(t)}$:
\begin{equation}\label{ZIuseful}
    \Z_I(\{s_i\},t) = \bra{1} T(\{s_i\}) \ket{\rho(t)}\,.
\end{equation}
By plugging \eqref{ZIuseful} in \eqref{expfromZ}, one obtains the expression
\begin{equation}
    \vev{\I^{i_1}(t)\cdots \I^{i_n}(t)} = \bra{1} P_{\I}^{i_{1}} \cdots P_{\I}^{i_{n}} \ket{\rho(t)} \,.
\end{equation}
Specifically, the probability for individual nodes $i$ to be infected at time $t$ is \eqref{expvalue}.
From the dynamical partition function \eqref{ZIuseful} it also follows that all higher moments of $I^i(t)$ are equal to the expectation value $\vev{\I^i(t)}$, since:
\begin{align}
    \vev{ (\I^i)^n (t) } & = \partial_{s_i}^n \Z(\{s_i\} ,t) |_{\{s_i\} = 0} \nonumber \\ 
    & = \bra{1} (P_\I^i)^n \ket{\rho(t)} = \bra{1} P_\I^i \ket{\rho(t)}\,.
\end{align}
Here we have used the fact that the projection operator is idempotent, i.e. $P_\I^2 = P_\I$. This implies that knowledge of the single node expectation values is sufficient to compute all higher-moments for the single nodes. Hence, the variance for individual nodes is:
\begin{equation}\label{varIi}
    {\rm Var}(\I^i(t) ) = \vev{\I^i(t)} - \vev{\I^i(t)}^2\,.
\end{equation}
For this reason we will focus in the remainder of the text on computing the single node expectation values \eqref{expvalue}. 

The expectation value for the prevalence in the whole network $\vev{\I(t)}$ is obtained by averaging \eqref{expvalue} over all nodes
\begin{equation}\label{prevalence}
    \vev{\I(t)} =  \frac{1}{N} \sum_{i}^N \vev{\I^i(t)}\,.
\end{equation}
This can be obtained from a dynamical partition function $\Z_I(s,t)$ by setting $s_i = s/N$ for all $i$ in \eqref{ZI}:
\begin{align}\label{ZItot}
    \Z_I(s,t) &= \bra{1} \exp\left(\frac{s}{N} \sum_{i=1}^N P_\I^i  \right) \ket{\rho(t)}\,, 
    \\
    \partial_s \Z_I(s,t) |_{s=0} &= \frac{1}{N} \sum_{i=1}^N \bra{1}P_{\I}^i \ket{\rho(t)} =  \vev{\I(t)}
\end{align}
By using the link between \eqref{ZI} and \eqref{ZItot}, it is also possible to express the variance of the prevalence in the ensemble average over trajectories as:
\begin{equation}
\label{eq:variance}
    {\rm Var}(\I(t)) = \frac{1}{N}\vev{\I(t)} + \frac{1}{N^2} \sum_{\substack{i,j = 1\\ i \neq j}}^N \bra{1} P_\I^i P_\I^j \ket{\rho(t)} - \vev{\I(t)}^2  \,.
\end{equation}
Unlike with the variance for individual nodes \eqref{varIi}, the total variance depends on the joint probability for nodes $i$ and $j$ to be infected simultaneously $\vev{\I^i\I^j(t)} = \bra{1} P_\I^i P_\I^j \ket{\rho(t)} $ for all node pairs. In the diagrammatic expansion, these terms receive contributions from diagrams where both $i$ and $j$ are infinitesimal stochastic nodes (sinks). Higher-order moments are also obtainable from this formalism, but in this paper we will focus only on single node expectation values.

\section{Diagrammatic expansion}\label{sec:diagrams}
We will now present a systematic procedure for computing the single node expectation values \eqref{expvalue}. Its expression depends on the time-dependent probability vector $\ket{\rho(t)}$. By the master equation \eqref{mastereqn} we can express it as  $\ket{\rho(t)} = \exp(t \Q) \ket{\mathcal{C}_0}$, where the initial state vector $\ket{\mathcal{C}_0}$ corresponds to the initial configuration of sources and susceptible nodes. The probability vector can be expanded as:
\begin{equation}\label{rhoexp}
\ket{\rho(t)} = \left( \mathbb{1} + t\,  \Q + \frac{t^2}{2!} \Q^2 + \frac{t^3}{3!} \Q^3 +  \ldots \right) \ket{\mathcal{C}_0} \,.
\end{equation}
Here henceforth, we adopt a graphical notation to represent the action of the operators $P_{\I}^i$, $\qSI^j$ appearing in the infinitesimal generator $\Q(A)$. We suppress displaying any susceptible nodes, as they have not been acted upon by either of the operators; we denote infected nodes as white circles, and sinks as blue ones: 
\begin{equation}
    \ket{\I} =\begin{array}{c}
		\begin{tikzpicture}
		\node[shape=circle,draw=black] (A) at (0,0) {};
		\end{tikzpicture}
		\end{array} \,, \qquad 
	\ket{\I} - \ket{\S} =
		\begin{array}{c}
		\begin{tikzpicture}
		\node[shape=circle,draw=black,fill=blue!20] (A) at (0,0) {};
		\end{tikzpicture}
		\end{array} \,.
\end{equation}
Any source node will be displayed as a crossed-out white node: ${\scriptsize \begin{array}{c}
		\begin{tikzpicture}
        \node[cross out,draw=black] (x) at (0,0) {};
		\node[shape=circle,draw=black] (A) at (0,0) {};
		\end{tikzpicture}
		\end{array}}$.
The local operators $P_\I$ and $\qSI$ in \eqref{operatordef} act on the colored and susceptible nodes as:
\begin{align}\label{operatorrules}
	P_\I\ket{\S} & = 0 \,, &
	P_\I \begin{array}{c}
	\begin{tikzpicture}
	\node[shape=circle,draw=black] (A) at (0,0) {};
	\end{tikzpicture}
	\end{array} & = 
	\begin{array}{c}
	\begin{tikzpicture}
	\node[shape=circle,draw=black] (A) at (0,0) {};
	\end{tikzpicture}
	\end{array}\,, &
	P_\I 			\begin{array}{c}
	\begin{tikzpicture}
	\node[shape=circle,draw=black,fill=blue!20] (A) at (0,0) {};
	\end{tikzpicture}
	\end{array} & =
	\begin{array}{c}
	\begin{tikzpicture}
	\node[shape=circle,draw=black] (A) at (0,0) {};
	\end{tikzpicture}
	\end{array} \,, \\ \nonumber
    \qSI \ket{\S} & = 
    \begin{array}{c}
    \begin{tikzpicture}
    \node[shape=circle,draw=black,fill=blue!20] (A) at (0,0) {};
    \end{tikzpicture}
    \end{array}\,,  &
    \qSI \begin{array}{c}
	\begin{tikzpicture}
	\node[shape=circle,draw=black] (A) at (0,0) {};
	\end{tikzpicture}
	\end{array} & = 0 \,,
	 &
	\qSI \begin{array}{c}
	\begin{tikzpicture}
	\node[shape=circle,draw=black,fill=blue!20] (A) at (0,0) {};
	\end{tikzpicture}
	\end{array} & = -
	\begin{array}{c}
	\begin{tikzpicture}
	\node[shape=circle,draw=black,fill=blue!20] (A) at (0,0) {};
	\end{tikzpicture}
	\end{array} \,.
\end{align}
In addition, the operator $\Q$ adds a directed edge pointing from $i$ to $j$ as specified by the adjacency matrix $A^{ij}$. The expansion \eqref{rhoexp} can now be organized as a perturbative series of directed diagrams with an increasing number of edges. When starting from an initial configuration with a single source node, the most general diagrammatic expansion for subgraphs with up to three edges reads:
\begin{widetext}
\begin{align}\label{diagramsexp}
\ket{\rho(t)} =  \begin{array}{c}
\begin{tikzpicture}
\node[cross out,draw=black] (x) at (0,0) {};
\node[shape=circle,draw=black] (A) at (0,0) {};
\end{tikzpicture}
\end{array} + \sum_{\rm s.g.} \sum_{n=1}^{\infty} \frac{t^n}{n!} \Bigg(  &
{\scriptsize \begin{array}{c}
\begin{tikzpicture}
\node[cross out,draw=black] (x) at (0,0) {};
\node[shape=circle,draw=black] (A) at (0,0) {};
\node[shape=circle,draw=black,fill=blue!20] (B) at (.5,0) {};
\path [->] (A) edge node[left] {} (B);
\end{tikzpicture}
\end{array}} 
+ {\scriptsize 
\begin{array}{c}
\begin{tikzpicture}
\node[cross out,draw=black] (x) at (0,0) {};
\node[shape=circle,draw=black] (A) at (0,0) {};
\node[shape=circle,draw=black,fill=blue!20] (B) at (.5,-.25) {};
\node[shape=circle,draw=black,fill=blue!20] (C) at (.5,.25) {};
\path [->] (A) edge node[left] {} (B);
\path [->] (A) edge node[left] {} (C);
\end{tikzpicture}
\end{array}}
+ {\scriptsize 
\begin{array}{c}
\begin{tikzpicture}
\node[cross out,draw=black] (x) at (0,0) {};
\node[shape=circle,draw=black] (A) at (0,0) {};
\node[shape=circle,draw=black] (B) at (.5,0) {};
\node[shape=circle,draw=black,fill=blue!20] (C) at (1,0) {};
\path [->] (A) edge node[left] {} (B);
\path [->] (B) edge node[left] {} (C);
\end{tikzpicture}
\end{array}} 
+
{\scriptsize 
\begin{array}{c}
\begin{tikzpicture}
\node[cross out,draw=black] (x) at (0,0) {};
\node[shape=circle,draw=black] (A) at (0,0) {};
\node[shape=circle,draw=black,fill=blue!20] (B) at (.5,-.4) {};
\node[shape=circle,draw=black,fill=blue!20] (C) at (.5,.4) {};
\node[shape=circle,draw=black,fill=blue!20] (D) at (.5,0) {};
\path [->] (A) edge (B);
\path [->] (A) edge (C);
\path [->] (A) edge (D);
\end{tikzpicture}
\end{array}} 
+
{\scriptsize
\begin{array}{c}
\begin{tikzpicture}
\node[cross out,draw=black] (x) at (0,0) {};
\node[shape=circle,draw=black] (A) at (0,0) {};
\node[shape=circle,draw=black] (B) at (.5,0.3) {};
\node[shape=circle,draw=black,fill=blue!20] (C) at (.5,-.3) {};
\node[shape=circle,draw=black,fill=blue!20] (D) at (1,0.3) {};
\path [->] (A) edge node[left] {} (B);
\path [->] (A) edge node[left] {} (C);
\path [->] (B) edge (D);
\end{tikzpicture}
\end{array}} 
+
{\scriptsize 
\begin{array}{c}
\begin{tikzpicture}
\node[cross out,draw=black] (x) at (0,0) {};
\node[shape=circle,draw=black] (A) at (0,0) {};
\node[shape=circle,draw=black] (B) at (.5,0) {};
\node[shape=circle,draw=black,fill=blue!20] (C) at (1,-0.25) {};
\node[shape=circle,draw=black,fill=blue!20] (D) at (1,0.25) {};
\path [->] (A) edge node[left] {} (B);
\path [->] (B) edge node[left] {} (C);
\path [->] (B) edge node[left] {} (D);
\end{tikzpicture}
\end{array}} 
+
{\scriptsize 
\begin{array}{c}
\begin{tikzpicture}
\node[cross out,draw=black] (x) at (0,0) {};
\node[shape=circle,draw=black] (A) at (0,0) {};
\node[shape=circle,draw=black] (B) at (.5,0) {};
\node[shape=circle,draw=black] (C) at (1,0) {};
\node[shape=circle,draw=black,fill=blue!20] (D) at (1.5,0) {};
\path [->] (A) edge (B);
\path [->] (B) edge (C);
\path [->] (C) edge (D);
\end{tikzpicture}
\end{array}}   
+
{\scriptsize
\begin{array}{c}
\begin{tikzpicture}
\node[cross out,draw=black] (x) at (0,0) {};
\node[shape=circle,draw=black] (A) at (0,0) {};
\node[shape=circle,draw=black] (B) at (.5,-.3) {};
\node[shape=circle,draw=black,fill=blue!20] (C) at (.5,.3) {};
\path [->] (A) edge (B);
\path [->] (A) edge (C);
\path [->] (B) edge (C);
\end{tikzpicture}
\end{array}} 
+ \ldots \Bigg) \,.
\end{align}
\end{widetext}
Where the dots denote diagrams with more than three edges.
Here each diagram $G$ graphically represents a vector $a_G[n]\ket{G}$. The combinatorial coefficient $a_G[n]$ counts the number of ways $G$ can be constructed by $n$ applications of $\Q$ on the initial configuration, up to a sign discussed below. The vector $\ket{G}$ is created as the tensor product of the displayed nodes states, times all the suppressed susceptibles $\ket{\S}$. Subgraph diagrams of the displayed topology can appear in the network $A$ with a multiplicity, which is accounted for by the sum over subgraphs (s.g.) in the above equation. Depending on the network $A$ some of the diagrams might not appear in the expansion, i.e. their multiplicity is zero.

\section{Dynamics of the SI system.} 
\label{sec:dynamics}
\begin{figure}
	\includegraphics[width=\columnwidth, frame, trim= {1.8cm .4cm 2.2cm .4cm}, clip]{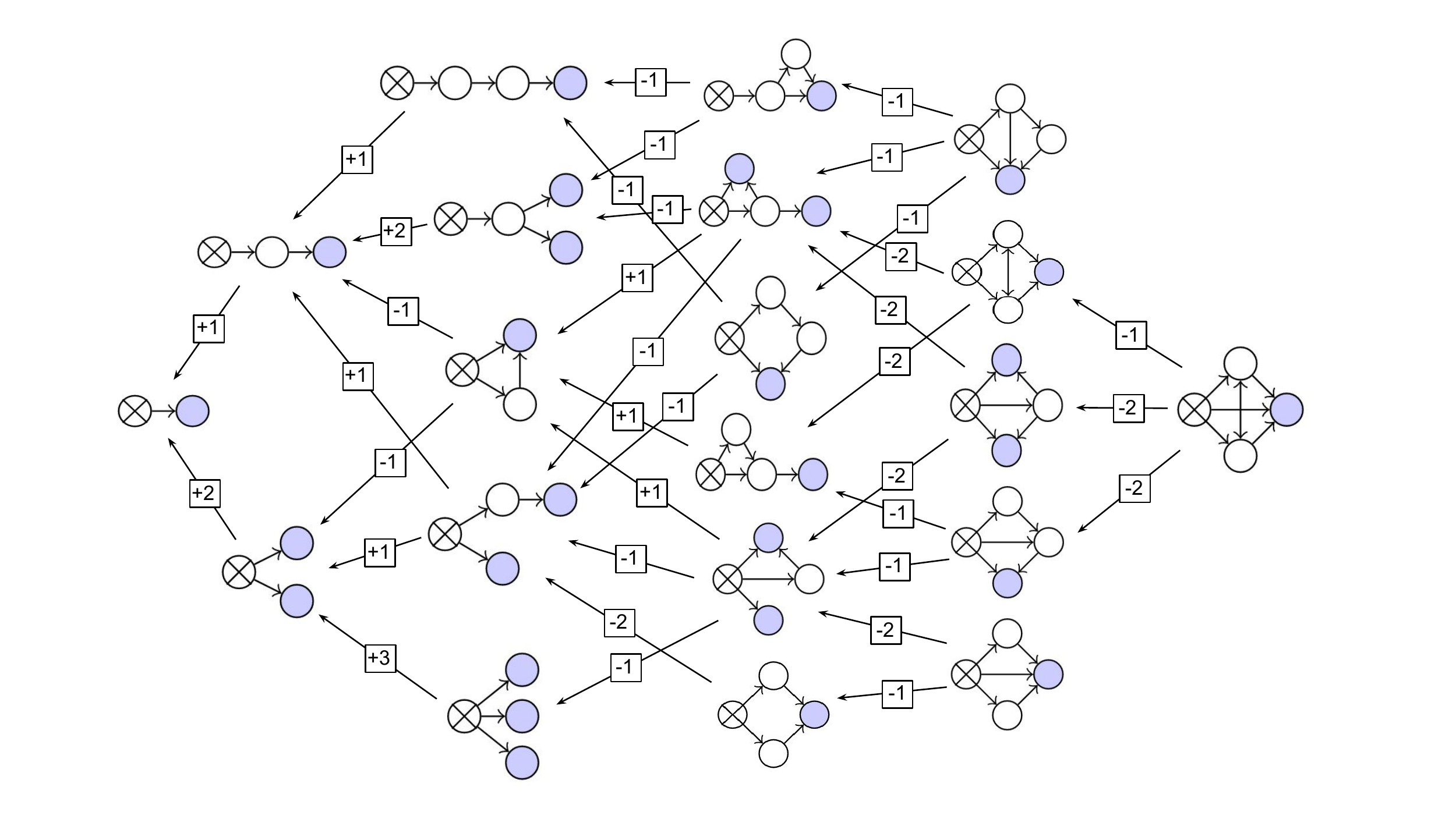}
	\caption{\label{fig:diagrams} The graph of parental relations between diagrams with up to four nodes. Each diagram contribution as a function of $t$ can be computed by performing the integral \eqref{diagramintegral}. The parental contribution $p_G(t)$ for each diagram is obtained by the sum of the parent diagrams times edge weights, $\alpha_{G H} $, denoted inside the squares. }
\end{figure}

To compute the sum over $n$ in \eqref{diagramsexp} we must obtain the coefficients $a_G[n]$ for each diagram. To solve this combinatorial problem we consider the action of $\Q$ on an arbitrary diagrams state vector $\ket{G}$. This defines four rules from which diagrams with $m+1$ edges are constructed from `parent diagrams' with $m$ edges.
Two rules govern the addition of new nodes to the diagram: 
\begin{itemize}
    \item[Rule 1:] If $\qSI$ acts on a new susceptible node and $P_\I$ on a white node, a new sink is attached to the white node by a directed edge.
    \item[Rule 2:]  If $\qSI$ acts on a new susceptible node and $P_\I$ on a blue node, the latter node is turned white and a new sink is attached to it.
\end{itemize}
In addition, internal edges between non-susceptible nodes are added to the diagrams by two more rules: 
\begin{itemize}
    \item[Rule 3:] When $\qSI$ acts on a blue node and $P_\I$ on a white node, an edge is added from the latter to the former node and a minus sign appears 
    \item[Rule 4:] When $\qSI$ acts on a blue node and $P_\I$ acts on a blue node, this node is turned white and a minus sign appears
\end{itemize}
Diagrammatically, examples of these rules are:
\begin{align} \label{rule1}
	\text{Rule 1}: & & {\scriptsize 
	\begin{array}{c}
	\begin{tikzpicture}
	\node[shape=circle,draw=black] (A) at (0,0) {};
	\node[shape=circle,draw=black,fill=blue!20] (B) at (.5,0) {};
	\path [->] (A) edge node[left] {} (B);
	\end{tikzpicture}
	\end{array} } & \to {\scriptsize
	\begin{array}{c}
	\begin{tikzpicture}
	\node[shape=circle,draw=black] (A) at (0,0) {};
	\node[shape=circle,draw=black,fill=blue!20] (B) at (.5,-.3) {};
	\node[shape=circle,draw=black,fill=blue!20] (C) at (.5,.3) {};
	\path [->] (A) edge (B);
	\path [->] (A) edge (C);
	\end{tikzpicture}
	\end{array}  }\\
	\label{rule2} \text{Rule 2}: & & {\scriptsize
	\begin{array}{c}
	\begin{tikzpicture}
	\node[shape=circle,draw=black] (A) at (0,0) {};
	\node[shape=circle,draw=black,fill=blue!20] (B) at (.5,0) {};
	\path [->] (A) edge node[left] {} (B);
	\end{tikzpicture}
	\end{array}} & \to  {\scriptsize
	\begin{array}{c}
	\begin{tikzpicture}
	\node[shape=circle,draw=black] (A) at (0,0) {};
	\node[shape=circle,draw=black] (B) at (.5,0) {};
	\node[shape=circle,draw=black,fill=blue!20] (C) at (1,0) {};
	\path [->] (A) edge node[left] {} (B);
	\path [->] (B) edge node[left] {} (C);
	\end{tikzpicture}
	\end{array} } \\
	\label{rule3} \text{Rule 3}: & & {\scriptsize 
    \begin{array}{c}
    \begin{tikzpicture}
    \node[shape=circle,draw=black] (A) at (0,0) {};
    \node[shape=circle,draw=black] (B) at (.5,0) {};
    \node[shape=circle,draw=black,fill=blue!20] (C) at (1,0) {};
    \path [->] (A) edge node[left] {} (B);
    \path [->] (B) edge node[left] {} (C);
    \end{tikzpicture}
    \end{array}} &  \to - {\scriptsize
	\begin{array}{c}
	\begin{tikzpicture}
	\node[shape=circle,draw=black] (A) at (0,0) {};
	\node[shape=circle,draw=black] (B) at (.5,-.3) {};
	\node[shape=circle,draw=black,fill=blue!20] (C) at (.5,.3) {};
	\path [->] (A) edge (B);
	\path [->] (A) edge (C);
	\path [->] (B) edge (C);
	\end{tikzpicture}
	\end{array} }
	\\ \label{rule4} \text{Rule 4}: & & {\scriptsize
	\begin{array}{c}
	\begin{tikzpicture}
	\node[shape=circle,draw=black] (A) at (0,0) {};
	\node[shape=circle,draw=black,fill=blue!20] (B) at (.5,-.3) {};
	\node[shape=circle,draw=black,fill=blue!20] (C) at (.5,.3) {};
	\path [->] (A) edge node[left] {} (B);
	\path [->] (A) edge node[left] {} (C);
	\end{tikzpicture}
	\end{array} } & \to -  {\scriptsize
	\begin{array}{c}
	\begin{tikzpicture}
	\node[shape=circle,draw=black] (A) at (0,0) {};
	\node[shape=circle,draw=black] (B) at (.5,-.3) {};
	\node[shape=circle,draw=black,fill=blue!20] (C) at (.5,.3) {};
	\path [->] (A) edge (B);
	\path [->] (A) edge (C);
	\path [->] (B) edge (C);
	\end{tikzpicture}
	\end{array} }
	\end{align}
When following the third rule, the operators $P_\I$ and $\qSI$ could act on a node pair which are already connected. This operation does not change the number of edges in the diagram, only the sign of the coefficient. It implies that $a_G[n]$ receives a contribution $- c \, a_G[n-1]$, where $c$ is the number of edges ending in sinks. Together with the contributions from parent diagrams $p_G[n-1]$ we can derive a recursive relation for $a_G[n]$:
\begin{equation}\label{recrelation}
    a_G[n] = - c\, a_G[n-1] + p_G[n-1]\,,
\end{equation}
with
\begin{equation}\label{parentdef}
p_G[n] = \sum_{H \in P(G)} \alpha_{G H} a_H[n]\,.
\end{equation}
Here $P(G)$ denotes the set of parents of the graph $G$ and $\alpha_{G H}$ gives the multiplicity and sign of the corresponding parental relationship. 
The set $P(G)$ is obtained by tracing the rules \eqref{rule1}-\eqref{rule4} backwards, and contain only diagrams with one less edge ending a blue node.
The collection of all possible diagrams can be thought of as a weighted directed graph of diagrams, where edges represent the child$\to$parent relationships and the edge weights correspond to the $\alpha_{GH}$ in \eqref{parentdef}. We display this graph of diagrams explicitly for all possible diagrams with four nodes in figure \ref{fig:diagrams}. 

Equation \eqref{recrelation} can be solved iteratively. To this end, it is convenient to express each diagram as a function of $t$ by performing the sum over $n$ in \eqref{diagramsexp}. If we define 
\begin{equation}
    a_G(t) = \sum_{n=1}^{\infty} \frac{t^n}{n!} a_G[n]
\end{equation} for each diagram, the recursion relation \eqref{recrelation} becomes a differential equation
\begin{equation}\label{dyneqn}
    \frac{d}{dt} a_G(t) = -c\, a_G(t) + p_G(t)\,,
\end{equation}
with the initial condition $a_G(0)= 0$, except when $ G = {\scriptsize \begin{array}{c}
		\begin{tikzpicture}
		\node[cross out,draw=black] (x) at (0,0) {};
		\node[shape=circle,draw=black] (A) at (0,0) {};
		\end{tikzpicture}
\end{array}}$, for which $a_{\otimes}(0) = 1$. The solution for a diagram $G$ is given by
\begin{equation}\label{diagramintegral}
    a_G(t) = e^{-c\, t} \int_0^t e^{c\,s} p_G(s) \, {\rm d} s \,.
\end{equation}
So any diagrams contribution to the probability vector \eqref{diagramsexp} can be explicitly computed from knowledge of the contributions of parent diagrams. As the parent diagrams necessarily have one edge less, we can compute all contributions systematically by starting from the smallest possible diagram: a single source node. The expectation value $\vev{\I^i(t)}$ in \eqref{expvalue} is then obtained by summing the contributions $a_G(t)$ for all subgraphs $G$ with fixed source, where the node $i$ is the only sink in $G$.

\subsection{Simple example - $K_3$ network}
Let us present and work out a simple example: the complete graph with three nodes $K_3$. We suppose that one of the three nodes is the initially infected source node. There are then four terms in the expansion of \eqref{diagramsexp}:
\begin{equation}\label{K3contributions}
    {\scriptsize \begin{array}{c}
\begin{tikzpicture}
\node[cross out,draw=black] (x) at (0,0) {};
\node[shape=circle,draw=black] (A) at (0,0) {};
\node[shape=circle,draw=black,fill=blue!20] (B) at (.5,0) {};
\path [->] (A) edge node[left] {} (B);
\end{tikzpicture}
\end{array}}, \quad
{\scriptsize 
\begin{array}{c}
\begin{tikzpicture}
\node[cross out,draw=black] (x) at (0,0) {};
\node[shape=circle,draw=black] (A) at (0,0) {};
\node[shape=circle,draw=black,fill=blue!20] (B) at (.5,-.25) {};
\node[shape=circle,draw=black,fill=blue!20] (C) at (.5,.25) {};
\path [->] (A) edge node[left] {} (B);
\path [->] (A) edge node[left] {} (C);
\end{tikzpicture}
\end{array}}, \quad
{\scriptsize 
\begin{array}{c}
\begin{tikzpicture}
\node[cross out,draw=black] (x) at (0,0) {};
\node[shape=circle,draw=black] (A) at (0,0) {};
\node[shape=circle,draw=black] (B) at (.5,0) {};
\node[shape=circle,draw=black,fill=blue!20] (C) at (1,0) {};
\path [->] (A) edge node[left] {} (B);
\path [->] (B) edge node[left] {} (C);
\end{tikzpicture}
\end{array}}, \quad
{\scriptsize
\begin{array}{c}
\begin{tikzpicture}
\node[cross out,draw=black] (x) at (0,0) {};
\node[shape=circle,draw=black] (A) at (0,0) {};
\node[shape=circle,draw=black] (B) at (.5,-.3) {};
\node[shape=circle,draw=black,fill=blue!20] (C) at (.5,.3) {};
\path [->] (A) edge (B);
\path [->] (A) edge (C);
\path [->] (B) edge (C);
\end{tikzpicture}
\end{array}}\,.
\end{equation}
To obtain the contributions corresponding to each of these diagrams we can explicitly compute the integral \eqref{diagramintegral}, following the graph of parental relationships in figure \ref{fig:diagrams}. The first diagram only receives contributions from the single source node. We will denote its contribution as a function of $t$ by the diagram itself, appended by $(t)$:
\begin{equation}
    {\scriptsize \begin{array}{c}
\begin{tikzpicture}
\node[cross out,draw=black] (x) at (0,0) {};
\node[shape=circle,draw=black] (A) at (0,0) {};
\node[shape=circle,draw=black,fill=blue!20] (B) at (.5,0) {};
\path [->] (A) edge node[left] {} (B);
\end{tikzpicture}
\end{array}}(t) = e^{-t} \int_0^t e^{s} a_{\otimes}(s) ds = 1 -e^{-t}
\end{equation}
Here we have used the initial condition $a_{\otimes}(t) = 1$. The next two diagrams in \eqref{K3contributions} can be derived from this result as
\begin{align}
    {\scriptsize 
\begin{array}{c}
\begin{tikzpicture}
\node[cross out,draw=black] (x) at (0,0) {};
\node[shape=circle,draw=black] (A) at (0,0) {};
\node[shape=circle,draw=black,fill=blue!20] (B) at (.5,-.25) {};
\node[shape=circle,draw=black,fill=blue!20] (C) at (.5,.25) {};
\path [->] (A) edge node[left] {} (B);
\path [->] (A) edge node[left] {} (C);
\end{tikzpicture}
\end{array}}(t) & = e^{-2t} \int_0^t e^{2s} \, 2 (1- e^{-s}) ds \\ \nonumber & = (1-e^{-t})^2 \,, \\
{\scriptsize 
\begin{array}{c}
\begin{tikzpicture}
\node[cross out,draw=black] (x) at (0,0) {};
\node[shape=circle,draw=black] (A) at (0,0) {};
\node[shape=circle,draw=black] (B) at (.5,0) {};
\node[shape=circle,draw=black,fill=blue!20] (C) at (1,0) {};
\path [->] (A) edge node[left] {} (B);
\path [->] (B) edge node[left] {} (C);
\end{tikzpicture}
\end{array}}(t) & = e^{-t} \int_0^t e^{s} (1-e^{-s}) ds \nonumber \\ 
& = (1 - e^{-t}(t+1) ) \,.
\end{align}
The final diagram receive parent contributions from both these diagrams, with a relative sign due to \eqref{rule3} and \eqref{rule4}:
\begin{align}
{\scriptsize
\begin{array}{c}
\begin{tikzpicture}
\node[cross out,draw=black] (x) at (0,0) {};
\node[shape=circle,draw=black] (A) at (0,0) {};
\node[shape=circle,draw=black] (B) at (.5,-.3) {};
\node[shape=circle,draw=black,fill=blue!20] (C) at (.5,.3) {};
\path [->] (A) edge (B);
\path [->] (A) edge (C);
\path [->] (B) edge (C);
\end{tikzpicture}
\end{array}}(t) & = e^{-2t} \int_0^t e^{2s}\left(-  {\scriptsize 
\begin{array}{c}
\begin{tikzpicture}
\node[cross out,draw=black] (x) at (0,0) {};
\node[shape=circle,draw=black] (A) at (0,0) {};
\node[shape=circle,draw=black,fill=blue!20] (B) at (.5,-.25) {};
\node[shape=circle,draw=black,fill=blue!20] (C) at (.5,.25) {};
\path [->] (A) edge node[left] {} (B);
\path [->] (A) edge node[left] {} (C);
\end{tikzpicture}
\end{array}}(s) - {\scriptsize 
\begin{array}{c}
\begin{tikzpicture}
\node[cross out,draw=black] (x) at (0,0) {};
\node[shape=circle,draw=black] (A) at (0,0) {};
\node[shape=circle,draw=black] (B) at (.5,0) {};
\node[shape=circle,draw=black,fill=blue!20] (C) at (1,0) {};
\path [->] (A) edge node[left] {} (B);
\path [->] (B) edge node[left] {} (C);
\end{tikzpicture}
\end{array}}(s) \right) \nonumber \\
& = - (1- e^{-t})(1 - e^{-t}(t+1) )
\end{align}
The infected expectation value for each of the two initially susceptible nodes receives contributions only from subgraphs where the node of interest is blue. This can now be computed from the above contributions as:
\begin{align}\label{K3ex} \nonumber
\vev{\I^i(t)} & = {\scriptsize \begin{array}{c}
\begin{tikzpicture}
\node[cross out,draw=black] (x) at (0,0) {};
\node[shape=circle,draw=black] (A) at (0,0) {};
\node[shape=circle,draw=black,fill=blue!20] (B) at (.5,0) {};
\path [->] (A) edge node[left] {} (B);
\end{tikzpicture}
\end{array}}(t)  + {\scriptsize 
\begin{array}{c}
\begin{tikzpicture}
\node[cross out,draw=black] (x) at (0,0) {};
\node[shape=circle,draw=black] (A) at (0,0) {};
\node[shape=circle,draw=black] (B) at (.5,0) {};
\node[shape=circle,draw=black,fill=blue!20] (C) at (1,0) {};
\path [->] (A) edge node[left] {} (B);
\path [->] (B) edge node[left] {} (C);
\end{tikzpicture}
\end{array}}(t) +
{\scriptsize
\begin{array}{c}
\begin{tikzpicture}
\node[cross out,draw=black] (x) at (0,0) {};
\node[shape=circle,draw=black] (A) at (0,0) {};
\node[shape=circle,draw=black] (B) at (.5,-.3) {};
\node[shape=circle,draw=black,fill=blue!20] (C) at (.5,.3) {};
\path [->] (A) edge (B);
\path [->] (A) edge (C);
\path [->] (B) edge (C);
\end{tikzpicture}
\end{array}}(t) \\
& = 1 - e^{-2t}(1+t) 
\end{align}
While the loop contribution is a strictly negative function, it is balanced by two positive contributions corresponding to the two ways to reach the sink node in the loop diagram. 
Only the sum over all three diagrams corresponds to a probability and is hence bounded between $0$ and $1$. Using equations \eqref{prevalence} and \eqref{eq:variance} it is straightforward to compute the prevalence and variance for the $K_3$ network.

\subsection{Diagram properties}
For any diagram, $a_G(t)$ will be monotonically increasing (or decreasing) and converge to a positive (or negative) integer for $t \to \infty$. Hence, each diagram possesses a sign, which determines whether its contribution is strictly positive or negative. This sign can be derived by the following argument: consider a diagram $G$ with $n$ nodes (of which $n_i$ initial infected) and $m$ edges. The action of the generator $\Q(A)$ creates a directed edge between two nodes, $i\to j$ and may change their state through the action of the operators $P_\I$ and $\qSI$. 
The in-degree $k_i^{\rm in}$ for node $i$ counts the number of times this node has been acted upon with $\qSI$. There is no sign change when $\qSI$ acts on a susceptible node, but after the node is turned blue any consequent action of $\qSI$ on the node produces a minus sign. If the node is ultimately projected upon by $P_\I^i$, no further sign changes occur. Hence the sign contribution from each white or blue node is $(-)^{k_i^{\rm in}-1}$. The total sign of the diagram $G$ is then the product over all blue and white nodes, which can be written as
\begin{equation}\label{sign}
    {\rm sign}(G) =\prod_{i=1}^{n-n_i} (-)^{k_i^{\rm in}-1}=(-)^{m-n+n_i}
\end{equation}
Here we have used the fact that initially infected nodes have no incoming edges, so the sum over all in-degrees of white and blue nodes equals the total number of edges in the diagram, $m$. 

In the above derivation, we are assuming that the subgraph $G$ has only single directed edges. For larger graphs, diagrams with double directed edges can appear, as there could be several paths from sources to sinks through a particular edge. The first example of a double directed diagram is:
\begin{equation}\label{excitation}
{\scriptsize 
	\begin{array}{c}
	\begin{tikzpicture}
	\node[cross out,draw=black] (x) at (0,0) {};
	\node[shape=circle,draw=black] (A) at (0,0) {};
	\node[shape=circle,draw=black,fill=blue!20] (B) at (1,0.3) {};
	\node[shape=circle,draw=black] (C) at (0.5,-0.3) {};
	\node[shape=circle,draw=black] (D) at (0.5,0.3) {};
	\node[shape=circle,draw=black,fill=blue!20] (E) at (1,-0.3) {};	
	\path [->] (A) edge node[left] {} (C);
	\path [->] (C) edge node[left] {} (E);
	\path [<-] (B) edge node[left] {} (D);
	\path [<->] (C) edge (D);
	\path [->] (A) edge (D);
	\end{tikzpicture}
	\end{array}}
\end{equation}
The double directed edge implies that some of the parents of this diagram have the arrow pointing one way, while other parents have it in the other direction. In this case, when computing the sign of the diagram, the double directed edges should be counted only once, and hence $m$ in \eqref{sign} represents the total number of edges of the undirected diagram. 

The expectation value $\vev{\I^i(t)}$ receives contributions from all subgraphs where node $i$ is the only sink. This sum is, by construction, bounded between 0 and 1, even though the individual terms may not be. In fact, the individual contributions $a_G(t)$ to $\ket{\rho(t)}$ do not correspond to a probability. Some diagrams have strictly positive contributions and others have strictly negative contributions. In appendix \ref{app:latetime} we prove that in the limit of infinite time, the diagrams contributions converge to a finite integer value $k_G$:
\begin{equation}\label{limitingvalue}
    k_{G}= {\rm sign}(G) \, n_o\,,
\end{equation} 
where $n_o$ denotes the number of valid orientations the double directed arrows in the diagram $G$ can take on. (For example, $n_o =2$ for \eqref{excitation} as the double directed arrow can point both up or down.)
Since \eqref{limitingvalue} can take on any integer value, it is clear that only the final sum over all contributing diagrams is bounded as a probability; the individual contributions will conspire to balance negative contributions with positive ones.

The number of subgraphs contributing to the expansion \eqref{diagramsexp} grows quickly with the number of edges. However, not all diagrams will give independent and new contributions. We have found and proven a set of symmetry relations and decomposition rules on the diagrams which we will now discuss.

\section{Symmetry relations}
The dynamics of the $\SI$ system gives rise to a number of relations between the contributions of different diagrams. For notational simplicity, each diagram below will now immediately represent its function $a_G(t)$. 
\subsection{Reverse the flow: sources $\leftrightarrow$ sinks}
The first relation states that $a_G(t)$ is equal, up to a sign, to $a_{G'}(t)$, where $G'$ is obtained from $G$ by transforming all of its $n_i$ sources into sinks and all of its $n_b$ sinks into sources. 
\begin{equation}\label{swapXB}
    {\scriptsize 
	\begin{array}{c}
	\begin{tikzpicture}
	\node[cross out,draw=black] (x) at (0,.5) {};
	\node[shape=circle,draw=black] (A) at (0,.5) {};
	\node[cross out,draw=black] (y) at (0,-.5) {};
	\node[shape=circle,draw=black] (A2) at (0,-.5) {};
	\node[shape=circle] (E) at (-.2,0.1) {$n_i \, \vdots $};
	\node[shape=circle,draw=black] (B) at (.5,0) {$\overrightarrow{s.g.}$};
	\node[shape=circle,draw=black,fill=blue!20] (C) at (1,-0.5) {};
	\node[shape=circle,draw=black,fill=blue!20] (D) at (1,0.5) {};
	\node[shape=circle] (E) at (1.2,0.1) {$\vdots  \, n_b $};
	\path [->] (A) edge node[left] {} (B);
	\path [->] (A2) edge node[left] {} (B);
	\path [->] (B) edge node[left] {} (C);
	\path [->] (B) edge node[left] {} (D);
	\end{tikzpicture}
	\end{array}}  = (-1)^{n_i-n_b} 
    {\scriptsize 
	\begin{array}{c}
	\begin{tikzpicture}
	\node[cross out,draw=black] (x) at (1,.5) {};
	\node[shape=circle,draw=black,fill=blue!20] (A) at (0,.5) {};
	\node[cross out,draw=black] (y) at (1,-.5) {};
	\node[shape=circle,draw=black,fill=blue!20] (A2) at (0,-.5) {};
	\node[shape=circle] (E) at (-.2,0.1) {$n_i \, \vdots $};
	\node[shape=circle,draw=black] (B) at (.5,0) {$\overleftarrow{s.g.}$};
	\node[shape=circle,draw=black] (C) at (1,-0.5) {};
	\node[shape=circle,draw=black] (D) at (1,0.5) {};
	\node[shape=circle] (E) at (1.2,0.1) {$\vdots  \, n_b $};
	\path [<-] (A) edge node[left] {} (B);
	\path [<-] (A2) edge node[left] {} (B);
	\path [<-] (B) edge node[left] {} (C);
	\path [<-] (B) edge node[left] {} (D);
	\end{tikzpicture}
	\end{array}}  \,,
\end{equation}
Here $\overrightarrow{s.g.}$ denotes any arbitrary directed subgraph consisting solely out of white nodes. 
This relation states that the flow of information from source to sink is invariant under changing the direction of the flow, up to a possible sign.

Some of the first non-trivial instances of this relation are
\begin{align}\label{swappingBX}
    {\scriptsize 
	\begin{array}{c}
	\begin{tikzpicture}
	\node[cross out,draw=black] (x) at (0,0) {};
	\node[shape=circle,draw=black] (A) at (0,0) {};
	\node[shape=circle,draw=black] (B) at (.5,0) {};
	\node[shape=circle,draw=black,fill=blue!20] (C) at (1,0) {};
	\node[shape=circle,draw=black] (D) at (0.25,0.4) {};
	\path [->] (A) edge node[left] {} (B);
	\path [->] (B) edge node[left] {} (C);
	\path [<-] (B) edge node[left] {} (D);
	\path [<-] (D) edge (A);
	\end{tikzpicture}
	\end{array}}(t) =
	{\scriptsize 
	\begin{array}{c}
	\begin{tikzpicture}
	\node[cross out,draw=black] (x) at (0,0) {};
	\node[shape=circle,draw=black] (A) at (0,0) {};
	\node[shape=circle,draw=black] (B) at (.5,0) {};
	\node[shape=circle,draw=black,fill=blue!20] (C) at (1,0) {};
	\node[shape=circle,draw=black] (D) at (0.75,0.4) {};
	\path [<-] (C) edge node[left] {} (D);
	\path [->] (B) edge node[left] {} (C);
	\path [->] (B) edge node[left] {} (D);
	\path [<-] (B) edge (A);
	\end{tikzpicture}
	\end{array}}(t)\,,
\end{align}
\qquad and:
\begin{equation}
    {\scriptsize 
	\begin{array}{c}
	\begin{tikzpicture}
	\node[cross out,draw=black] (x) at (0,0) {};
	\node[shape=circle,draw=black] (A) at (0,0) {};
	\node[shape=circle,draw=black] (B) at (.5,0) {};
	\node[shape=circle,draw=black,fill=blue!20] (C) at (1,-0.25) {};
	\node[shape=circle,draw=black,fill=blue!20] (D) at (1,0.25) {};
	\path [->] (A) edge node[left] {} (B);
	\path [->] (B) edge node[left] {} (C);
	\path [->] (B) edge node[left] {} (D);
	\end{tikzpicture}
	\end{array}}(t)  = -
	{\scriptsize 
	\begin{array}{c}
	\begin{tikzpicture}
	\node[cross out,draw=black] (y) at (0,0.25) {};
	\node[cross out,draw=black] (x) at (0,-.25) {};
	\node[shape=circle,draw=black] (A) at (0,0.25) {};
	\node[shape=circle,draw=black] (B) at (.5,0) {};
	\node[shape=circle,draw=black] (C) at (0,-0.25) {};
	\node[shape=circle,draw=black,fill=blue!20] (D) at (1,0) {};
	\path [->] (A) edge node[left] {} (B);
	\path [<-] (B) edge node[left] {} (C);
	\path [->] (B) edge node[left] {} (D);
	\end{tikzpicture}
	\end{array}}(t) \,,
\end{equation}
The invariance under reversal of the flow of the diagram can be understood from the way the diagrams are constructed. For each edge in the diagram, a single operation of $A^{ij}P_\I^i \qSI^j$ must have been applied to construct the edge from $i$ to $j$ \footnote{We assume the underlying network topology has only undirected edges, such that $A^{ij} = A^{ji}$}. If we replace
${\scriptsize \begin{array}{c}
		\begin{tikzpicture}
        \node[cross out,draw=black] (x) at (0,0) {};
		\node[shape=circle,draw=black] (A) at (0,0) {};
		\end{tikzpicture}
		\end{array}} \leftrightarrow 
{\scriptsize \begin{array}{c}
		\begin{tikzpicture}
		\node[shape=circle,draw=black,fill=blue!20] (A) at (0,0) {};
		\end{tikzpicture}
\end{array}}$, 
then the diagram of the same topology is constructed by applying $A^{ij} P_\I^j \qSI^i$ to the node pairs $(i,j)$, creating the same edges in the opposite direction. Since the total number of operations to create this inverted diagram is unchanged, the only difference between the two contributions can be a sign. The sign arises because to obtain the flipped diagram the operator $\qSI$ may act on a different number of blue nodes. Specifically, from \eqref{sign}, by reversing the flow the contribution gets a minus sign if the difference $(n_i-n_b)$ is odd.

An extension of the above rule exists for diagrams whose only parents are related by the above symmetry relation. For instance, since the parents of the first two diagrams below satisfy the relation \eqref{swappingBX}, it follows that:
\begin{equation}
    	{\scriptsize 
	\begin{array}{c}
	\begin{tikzpicture}
	\node[cross out,draw=black] (x) at (0,0) {};
	\node[shape=circle,draw=black] (A) at (0,0) {};
	\node[shape=circle,draw=black] (B) at (.5,0) {};
	\node[shape=circle,draw=black] (C) at (1,0) {};
	\node[shape=circle,draw=black] (D) at (0.75,0.4) {};
	\node[shape=circle,draw=black,fill=blue!20] (E) at (1.5,0) {};
	\path [<-] (C) edge node[left] {} (D);
	\path [->] (B) edge node[left] {} (C);
	\path [->] (B) edge node[left] {} (D);
	\path [<-] (B) edge (A);
	\path[->] (C) edge (E) ;
	\end{tikzpicture}
	\end{array}}(t) =
	{\scriptsize 
	\begin{array}{c}
	\begin{tikzpicture}
	\node[cross out,draw=black] (x) at (0,0) {};
	\node[shape=circle,draw=black] (A) at (0,0) {};
	\node[shape=circle,draw=black] (B) at (.5,0) {};
	\node[shape=circle,draw=black] (C) at (1,0) {};
	\node[shape=circle,draw=black] (D) at (0.25,0.4) {};
	\node[shape=circle,draw=black,fill=blue!20] (E) at (1.5,0) {};
	\path [<-] (B) edge node[left] {} (D);
	\path [->] (B) edge node[left] {} (C);
	\path [->] (A) edge node[left] {} (D);
	\path [<-] (B) edge (A);
	\path[->] (C) edge (E) ;
	\end{tikzpicture}
	\end{array}}(t) =
	{\scriptsize 
	\begin{array}{c}
	\begin{tikzpicture}
	\node[cross out,draw=black] (x) at (0,0) {};
	\node[shape=circle,draw=black] (A) at (0,0) {};
	\node[shape=circle,draw=black] (B) at (.5,0) {};
	\node[shape=circle,draw=black] (C) at (1,0) {};
	\node[shape=circle,draw=black] (D) at (1.25,0.4) {};
	\node[shape=circle,draw=black,fill=blue!20] (E) at (1.5,0) {};
	\path [->] (C) edge node[left] {} (D);
	\path[->]  (C) edge (E);
	\path [->] (B) edge node[left] {} (C);
	\path [<-] (B) edge (A);
	\path[->] (D) edge (E) ;
	\end{tikzpicture}
	\end{array}}(t)\,,
\end{equation}
where the last diagram is obtained from the second by again reversing the flow.
This relation generalizes to any diagram consisting out of a $n$ node chain with an arbitrary subgraph $g$ in the middle. For any $k \leq n$, we have diagrammatically:
\begin{align}
& \quad {\scriptsize
	\begin{array}{c}
	\begin{tikzpicture}
	\node[cross out,draw=black] (x) at (0,0) {};
	\node[shape=circle,draw=black] (A) at (0,0) {};
	\node[shape=circle] (dots) at (0.75,0) {$\dots$} ;
	\node[shape=circle] (A1) at (0.75,0.2) {$n-k$};
	\node[shape=circle,draw=black] (B) at (1.5,0) {};
	\node[shape=circle,draw=black] (b) at (3.5,0) {};
	\node[shape=circle] (dots2) at (4.25,0) {$\dots$} ;
	\node[shape=circle] (A2) at (4.25,0.2) {$k$};
	\node[shape=circle,draw=black,fill=blue!20] (G) at (5,0) {};
	\node[shape=circle] (E) at (3.1,0.1) {$\vdots $};
	\node[shape=circle] (C) at (2.6,0.52) {};
	\node[shape=circle] (D) at (2.6,-0.52) {};
	\node[shape=circle] (vdots) at (1.9,0.1) {$\vdots $};
	\node[shape=circle] (gn1) at (2.4,0.52) {};
	\node[shape=circle] (gn2) at (2.4,-0.52) {};
	\node[shape=circle,draw=black,minimum size=1cm,fill=white] (S) at (2.5,0) {$g$};
	\path [->] (C) edge node[left] {} (b);
    \path [->] (A) edge node[left] {} (dots);
    \path [->] (dots) edge node[left] {} (B);
    \path [->] (B) edge node[left] {} (gn1);
    \path [->] (B) edge node[left] {} (gn2);
	\path [->] (D) edge node[left] {} (b);
	\path [->] (b) edge node[left] {} (dots2);
	\path [->] (dots2) edge node[left] {} (G);
	\end{tikzpicture}
	\end{array}}
	\\ \nonumber
&	=  {\scriptsize
	\begin{array}{c}
	\begin{tikzpicture}
	\node[cross out,draw=black] (x) at (0,0) {};
	\node[shape=circle,draw=black] (A) at (0,0) {};
	\node[shape=circle] (dots) at (0.75,0) {$\dots$} ;
	\node[shape=circle,draw=black] (B) at (1.5,0) {};
	\node[shape=circle,draw=black,fill=blue!20] (b) at (3.5,0) {};
	\node[shape=circle] (A1) at (0.75,0.2) {$n$};
	\node[shape=circle] (E) at (3.1,0.1) {$\vdots $};
	\node[shape=circle] (C) at (2.6,0.52) {};
	\node[shape=circle] (D) at (2.6,-0.52) {};
	\node[shape=circle] (vdots) at (1.9,0.1) {$\vdots $};
	\node[shape=circle] (gn1) at (2.4,0.52) {};
	\node[shape=circle] (gn2) at (2.4,-0.52) {};
	\node[shape=circle,draw=black,minimum size=1cm,fill=white] (S) at (2.5,0) {$g$};
	\path [->] (C) edge node[left] {} (b);
    \path [->] (A) edge node[left] {} (dots);
    \path [->] (dots) edge node[left] {} (B);
    \path [->] (B) edge node[left] {} (gn1);
    \path [->] (B) edge node[left] {} (gn2);
	\path [->] (D) edge node[left] {} (b);
	\end{tikzpicture}
	\end{array}}  = 
    {\scriptsize
	\begin{array}{c}
	\begin{tikzpicture}
	\node[cross out,draw=black] (x) at (0,0) {};
	\node[shape=circle,draw=black] (A) at (0,0) {};
	\node[shape=circle,draw=black] (B) at (2,0) {};
	\node[shape=circle] (dots) at (2.75,0) {$\dots$} ;
	\node[shape=circle,draw=black,fill=blue!20] (b) at (3.5,0) {};
	\node[shape=circle] (A1) at (2.75,0.2) {$n$};
	\node[shape=circle] (E) at (1.6,0.1) {$\vdots $};
	\node[shape=circle] (C) at (1.1,0.52) {};
	\node[shape=circle] (D) at (1.1,-0.52) {};
	\node[shape=circle] (vdots) at (.4,0.1) {$\vdots $};
	\node[shape=circle] (gn1) at (.9,0.52) {};
	\node[shape=circle] (gn2) at (.9,-0.52) {};
	\node[shape=circle,draw=black,minimum size=1cm,fill=white] (S) at (1,0) {$g$};
	\path [->] (C) edge node[left] {} (B);
    \path [->] (A) edge node[left] {} (gn1);
    \path [->] (dots) edge node[left] {} (b);
    \path [->] (B) edge node[left] {} (dots);
    \path [->] (A) edge node[left] {} (gn2);
	\path [->] (D) edge node[left] {} (B);
	\end{tikzpicture}
	\end{array}}
\end{align}
Here $g$ represents an arbitrary graph consisting solely out of white nodes.

\subsection{Merging/separating sources and sinks}
The second relation states that, for a graph $G$ with $s$ sources of degree one, each connected to distinct nodes,  $a_G(t) = a_{G'}(t)$  where $G'$ is obtained from $G$ by merging all sources into a single source of degree $s$:
\begin{align}\label{mergingsources}
{\scriptsize
	\begin{array}{c}
	\begin{tikzpicture}
	\node[cross out,draw=black] (x) at (0,0.5) {};
	\node[shape=circle,draw=black] (A) at (0,0.5) {};
	\node[cross out,draw=black] (y) at (0,-0.5) {};
	\node[shape=circle,draw=black] (B) at (0,-0.5) {};
	\node[shape=circle] (s) at (-0.2,0) {$s$};
	\node[shape=circle] (E) at (0,0.1) {$\vdots $};
	\node[shape=circle,draw=black,minimum size=1cm] (S) at (0.9,0) {$g$};
	\path [->] (A) edge node[left] {} (S);
	\path [->] (B) edge node[left] {} (S);
	\end{tikzpicture}
	\end{array}}=&
	{\scriptsize
	\begin{array}{c}
	\begin{tikzpicture}
	\node[cross out,draw=black] (x) at (0,0) {};
	\node[shape=circle,draw=black] (A) at (0,0) {};
	\node[shape=circle] (C) at (0.85,0.5) {};
	\node[shape=circle] (D) at (0.85,-0.5) {};
	\node[shape=circle] (E) at (0.4,0.1) {$\vdots $};
	\node[shape=circle,draw=black,minimum size=1cm] (S) at (1,0) {$g$};
	\path [->] (A) edge node[left] {} (C);
	\path [->] (A) edge node[left] {} (D);
	\end{tikzpicture}
	\end{array}} \,.
\end{align}
By the first relation \eqref{swapXB} this implies that also $s$ sinks can be merged into one, now producing the sign $(-1)^{s-1}$:
\begin{align}
	 \label{mergingsinks}
	{\scriptsize
	\begin{array}{c}
	\begin{tikzpicture}
	\node[shape=circle,draw=black,fill=blue!20] (A) at (1,0.5) {};
	\node[shape=circle,draw=black,fill=blue!20] (B) at (1,-0.5) {};
	\node[shape=circle] (s) at (1.2,0) {$s$};
	\node[shape=circle] (E) at (1,0.1) {$\vdots $};
	\node[shape=circle,draw=black,minimum size=1cm] (S) at (0,0) {$\tilde{g}$};
	\path [->] (S) edge node[left] {} (A);
	\path [->] (S) edge node[left] {} (B);
	\end{tikzpicture}
	\end{array}}=&(-1)^{s-1}
	{\scriptsize
	\begin{array}{c}
	\begin{tikzpicture}
	\node[shape=circle,draw=black,fill=blue!20] (A) at (1,0) {};
	\node[shape=circle] (E) at (.6,0.1) {$\vdots $};
	\node[shape=circle] (C) at (0.1,0.52) {};
	\node[shape=circle] (D) at (0.1,-0.52) {};
	\node[shape=circle,draw=black,minimum size=1cm,fill=white] (S) at (0,0) {$\tilde{g}$};
	\path [->] (C) edge node[left] {} (A);
	\path [->] (D) edge node[left] {} (A);
	\end{tikzpicture}
	\end{array}}\,.
	\end{align}
In the above equalities, the subgraph $g$ can in principle contain other sources or white nodes, but it must contain at least one sink; contrarily, the subgraph $\tilde{g}$ must contain at least one source but it can contain other white or blue nodes. 

Two elementary examples of this rule are:
\begin{align}
{\scriptsize 
 	\begin{array}{c}
 	\begin{tikzpicture}
 	\node[cross out,draw=black] (x) at (0,0) {};
 	\node[shape=circle,draw=black] (A) at (0,0) {};
 	\node[shape=circle,draw=black] (B) at (.5,0) {};
 	\node[shape=circle,draw=black,fill=blue!20] (C) at (1,0) {};
 	\node[shape=circle,draw=black] (D) at (0.25,0.4) {};
 	\path [->] (A) edge node[left] {} (B);
 	\path [->] (B) edge node[left] {} (C);
 	\path [<-] (B) edge node[left] {} (D);
 	\path [<-] (D) edge (A);
 	\end{tikzpicture}
 	\end{array}}(t)
 & = {\scriptsize 
	\begin{array}{c}
	\begin{tikzpicture}
	\node[cross out,draw=black] (x) at (0,0) {};
	\node[cross out,draw=black] (y) at (0,0.4) {};
	\node[shape=circle,draw=black] (A) at (0,0) {};
	\node[shape=circle,draw=black] (E) at (0,0.4) {};
	\node[shape=circle,draw=black] (B) at (.5,0) {};
	\node[shape=circle,draw=black,fill=blue!20] (C) at (1,0) {};
	\node[shape=circle,draw=black] (D) at (0.5,0.4) {};
	\path [->] (A) edge node[left] {} (B);
	\path [->] (B) edge node[left] {} (C);
	\path [<-] (B) edge node[left] {} (D);
	\path [<-] (D) edge (E);
	\end{tikzpicture}
	\end{array}}(t)
 \,,  \\ 
{\scriptsize
	\begin{array}{c}
	\begin{tikzpicture}
	\node[cross out,draw=black] (x) at (0,0) {};
	\node[shape=circle,draw=black] (A) at (0,0) {};
	\node[shape=circle,draw=black,fill=blue!20] (B) at (.5,-.3) {};
	\node[shape=circle,draw=black] (C) at (.5,.3) {};
	\path [->] (A) edge (B);
	\path [->] (A) edge (C);
	\path [<-] (B) edge (C);
	\end{tikzpicture}
	\end{array}}(t)
 & = - {\scriptsize
	\begin{array}{c}
	\begin{tikzpicture}
	\node[cross out,draw=black] (x) at (0,0) {};
	\node[shape=circle,draw=black] (A) at (0,0) {};
	\node[shape=circle,draw=black] (B) at (.5,0.3) {};
	\node[shape=circle,draw=black,fill=blue!20] (C) at (.5,-.3) {};
	\node[shape=circle,draw=black,fill=blue!20] (D) at (1,0.3) {};
	\path [->] (A) edge node[left] {} (B);
	\path [->] (A) edge node[left] {} (C);
	\path [->] (B) edge (D);
	\end{tikzpicture}
	\end{array}}(t) 
\,.
\end{align}
A consequence of this relation is that diagrams containing multiple branches from the source node can be factorized into the product of the branches. For instance, the last diagram above can be further decomposed by separating the two branches from the source node, resulting in two disconnected diagrams. The diagram contribution thus factorizes into the contributions from the branches:
\begin{equation}
    {\scriptsize
	\begin{array}{c}
	\begin{tikzpicture}
	\node[cross out,draw=black] (x) at (0,0) {};
	\node[shape=circle,draw=black] (A) at (0,0) {};
	\node[shape=circle,draw=black] (B) at (.5,0.3) {};
	\node[shape=circle,draw=black,fill=blue!20] (C) at (.5,-.3) {};
	\node[shape=circle,draw=black,fill=blue!20] (D) at (1,0.3) {};
	\path [->] (A) edge node[left] {} (B);
	\path [->] (A) edge node[left] {} (C);
	\path [->] (B) edge (D);
	\end{tikzpicture}
	\end{array}}(t) 
	=
    {\scriptsize \begin{array}{c}
	\begin{tikzpicture}
	\node[cross out,draw=black] (x) at (0,0) {};
	\node[shape=circle,draw=black] (A) at (0,0) {};
	\node[shape=circle,draw=black,fill=blue!20] (B) at (.5,0) {};
	\path [->] (A) edge node[left] {} (B);
	\end{tikzpicture}
	\end{array}}(t) \times
{\scriptsize 
	\begin{array}{c}
	\begin{tikzpicture}
	\node[cross out,draw=black] (x) at (0,0) {};
	\node[shape=circle,draw=black] (A) at (0,0) {};
	\node[shape=circle,draw=black] (B) at (.5,0) {};
	\node[shape=circle,draw=black,fill=blue!20] (C) at (1,0) {};
	\path [->] (A) edge node[left] {} (B);
	\path [->] (B) edge node[left] {} (C);
	\end{tikzpicture}
	\end{array}}(t) \;
\end{equation}
In general, any diagram containing multiple branches from the source node factorizes into the product of the branches. Graphically:
\begin{equation}\label{productrule}
{\scriptsize
	\begin{array}{c}
	\begin{tikzpicture}
	\node[cross out,draw=black] (x) at (-1,0) {};
	\node[shape=circle,draw=black] (A) at (-1,0) {};
	\node[shape=circle](b1) at (.4,1){};
	\node[shape=circle](c1) at (.5,0.35){};
	\node[shape=circle](b2) at (.4,-1){};
	\node[shape=circle](c2) at (.4,-0.4){};
	\node[shape=circle] (e1) at (.08,0.52) {$\vphantom{\int\limits^x}\smash{\vdots}$};
	\node[shape=circle] (E) at (.55,0.1) {$\vdots $};
	\node[shape=circle] (e2) at (.08,-0.60) {$\vphantom{\int\limits^x}\smash{\vdots}$};
	\node[shape=circle,draw=black] (B1) at (.5,0.65) {$B_1$};
	\node[shape=circle,draw=black] (B2) at (.5,-0.65) {$B_n$};
	\path [->] (A) edge node[left] {} (b1);
	\path [->] (A) edge node[left] {} (c1);
	\path [->] (A) edge node[left] {} (b2);
	\path [->] (A) edge node[left] {} (c2);
	\end{tikzpicture}
	\end{array}} =
    {\scriptsize \begin{array}{c}
	\begin{tikzpicture}
	\node[cross out,draw=black] (x) at (0,0) {};
	\node[shape=circle,draw=black] (A) at (0,0) {};
	\node[shape=circle,draw=black] (B1) at (.9,0.0) {$B_1$};
	\node[shape=circle](b3) at (.85,0.35){};
	\node[shape=circle](c3) at (.83,-0.33){};
	\node[shape=circle] (e3) at (.5,0.0) {$\vphantom{\int\limits^x}\smash{\vdots}$};
	\path [->] (A) edge node[left] {} (b3);
	\path [->] (A) edge node[left] {} (c3);
	\end{tikzpicture}
	\end{array}}  \times \dots \times
{\scriptsize 
	\begin{array}{c}
	\begin{tikzpicture}
	\node[cross out,draw=black] (x) at (0,0) {};
	\node[shape=circle,draw=black] (A) at (0,0) {};
	\node[shape=circle,draw=black] (B1) at (.9,0.0) {$B_n$};
	\node[shape=circle](b3) at (.85,0.35){};
	\node[shape=circle](c3) at (.83,-0.33){};
	\node[shape=circle] (e3) at (.5,0.0) {$\vphantom{\int\limits^x}\smash{\vdots}$};
	\path [->] (A) edge node[left] {} (b3);
	\path [->] (A) edge node[left] {} (c3);
	\end{tikzpicture}
	\end{array}} \,.
\end{equation}
Here $B_1 , \ldots , B_n$ are $n$ mutually disconnected subgraphs containing at least one blue node. The proof follows immediately by separating the source into $n$ sources. As the spreading pathways on disconnected components correspond to independent events, the contribution factorizes into the product of the branches contributions. 
 
\subsection{Cutting off sources}\label{sec:cuttingsources}
A final relation involves diagrams with a single source of degree one. It is graphically:
\begin{equation}\label{cuttingsources}
    (c-1){\scriptsize
	\begin{array}{c}
	\begin{tikzpicture}
	\node[cross out,draw=black] (x) at (0,0) {};
	\node[shape=circle,draw=black] (O) at (0,0) {};
	\node[shape=circle,draw=black] (A) at (0.5,0) {};
	\node[shape=circle] (C) at (1.35,0.5) {};
	\node[shape=circle] (D) at (1.35,-0.5) {};
	\node[shape=circle] (E) at (.9,0.1) {$\vdots $};
	\node[shape=circle,draw=black,minimum size=1cm] (S) at (1.5,0) {$g$};
	\path [->] (O) edge node[left] {} (A);
	\path [->] (A) edge node[left] {} (C);
	\path [->] (A) edge node[left] {} (D);
	\end{tikzpicture}
	\end{array}}=
	{\scriptsize
	\begin{array}{c}
	\begin{tikzpicture}
	\node[cross out,draw=black] (x) at (0,0) {};
	\node[shape=circle,draw=black] (O) at (0,0) {};
	\node[shape=circle,draw=black] (A) at (0.5,0) {};
	\node[shape=circle] (C) at (1.35,0.5) {};
	\node[shape=circle] (D) at (1.35,-0.5) {};
	\node[shape=circle] (E) at (.9,0.1) {$\vdots $};
	\node[shape=circle,draw=black,minimum size=1cm] (S) at (1.5,0) {$p_g$};
	\path [->] (O) edge node[left] {} (A);
	\path [->] (A) edge node[left] {} (C);
	\path [->] (A) edge node[left] {} (D);
	\end{tikzpicture}
	\end{array}}
	-
	{\scriptsize
	\begin{array}{c}
	\begin{tikzpicture}
	\node[cross out,draw=black] (x) at (0,0) {};
	\node[shape=circle,draw=black] (A) at (0.,0) {};
	\node[shape=circle] (C) at (.85,0.5) {};
	\node[shape=circle] (D) at (.85,-0.5) {};
	\node[shape=circle] (E) at (.4,0.1) {$\vdots $};
	\node[shape=circle,draw=black,minimum size=1cm] (S) at (1,0) {$g$};
	\path [->] (A) edge node[left] {} (C);
	\path [->] (A) edge node[left] {} (D);
	\end{tikzpicture}
	\end{array}}
\end{equation}
Here $g$ represents a diagram with $c$ edges to sinks, and $p_g$ represents its parent contribution. The last diagram on the right hand side is obtained from the original diagram by moving the source node along its only edge. 
An explicit example of this rule is:
\begin{equation}\label{secbranch}
{\scriptsize 
	\begin{array}{c}
	\begin{tikzpicture}
	\node[cross out,draw=black] (x) at (0,0) {};
	\node[shape=circle,draw=black] (A) at (0,0) {};
	\node[shape=circle,draw=black] (B) at (.5,0) {};
	\node[shape=circle,draw=black,fill=blue!20] (C) at (1,-0.25) {};
	\node[shape=circle,draw=black,fill=blue!20] (D) at (1,0.25) {};
	\path [->] (A) edge node[left] {} (B);
	\path [->] (B) edge node[left] {} (C);
	\path [->] (B) edge node[left] {} (D);
	\end{tikzpicture}
	\end{array}}(t) = 2 {\scriptsize 
	\begin{array}{c}
	\begin{tikzpicture}
	\node[cross out,draw=black] (x) at (0,0) {};
	\node[shape=circle,draw=black] (A) at (0,0) {};
	\node[shape=circle,draw=black] (B) at (.5,0) {};
	\node[shape=circle,draw=black,fill=blue!20] (C) at (1,0) {};
	\path [->] (A) edge node[left] {} (B);
	\path [->] (B) edge node[left] {} (C);
	\end{tikzpicture}
	\end{array}}(t) -   {\scriptsize 
	\begin{array}{c}
	\begin{tikzpicture}
	\node[cross out,draw=black] (x) at (0,0) {};
	\node[shape=circle,draw=black] (A) at (0,0) {};
	\node[shape=circle,draw=black,fill=blue!20] (B) at (.5,-.25) {};
	\node[shape=circle,draw=black,fill=blue!20] (C) at (.5,.25) {};
	\path [->] (A) edge node[left] {} (B);
	\path [->] (A) edge node[left] {} (C);
	\end{tikzpicture}
	\end{array}}(t)
\end{equation}
In appendix \ref{app:cuttingproof} we provide a proof of this symmetry relationship. 

\subsection{Decomposing tree diagrams}
Using combinations of the symmetry relations, many diagrams containing a single source can be expressed in terms of (sums or products of) simpler diagrams. For instance, any tree diagram can be decomposed completely in terms of only chain diagrams by using \eqref{productrule} and \eqref{cuttingsources}. The contribution from chains of any length $d$ is given by \cite{merbis2021exact}:
\begin{equation}\label{Id}
\vev{\I^d(t)}_{\rm chain} = 1 - \frac{\Gamma(d,t)}{\Gamma(d)} \,,
\end{equation}
where $\Gamma(d,t)$ is the upper incomplete Gamma function
\begin{equation}\label{Gammadef}
\Gamma(d,t) = \int_t^\infty s^{d-1} e^{-s} ds = (d-1)! e^{-t} \sum_{n=0}^{d-1} \frac{t^n}{n!} \,.
\end{equation}
Any diagram which can be mapped to a tree by separating sinks can consequently also be decomposed in terms of chain diagrams. One example of such decomposition is:
\begin{align}
  & \quad  {\scriptsize 
	\begin{array}{c}
	\begin{tikzpicture}
	\node[cross out,draw=black] (x) at (0,0) {};
	\node[shape=circle,draw=black] (A) at (0,0) {};
	\node[shape=circle,draw=black] (B) at (.5,0) {};
	\node[shape=circle,draw=black,fill=blue!20] (C) at (1,0) {};
	\node[shape=circle,draw=black] (D) at (0.75,0.4) {};
	\path [<-] (C) edge node[left] {} (D);
	\path [->] (B) edge node[left] {} (C);
	\path [->] (B) edge node[left] {} (D);
	\path [<-] (B) edge (A);
	\end{tikzpicture}
	\end{array}}(t) = -
	{\scriptsize 
	\begin{array}{c}
	\begin{tikzpicture}
	\node[cross out,draw=black] (x) at (0,0) {};
	\node[shape=circle,draw=black] (A) at (0,0) {};
	\node[shape=circle,draw=black] (B) at (.5,0) {};
	\node[shape=circle,draw=black,fill=blue!20] (C) at (1,0) {};
	\node[shape=circle,draw=black] (D) at (0.5,0.4) {};
	\node[shape=circle,draw=black,fill=blue!20] (E) at (1,0.4) {};
	\path [<-] (E) edge node[left] {} (D);
	\path [->] (B) edge node[left] {} (C);
	\path [->] (B) edge node[left] {} (D);
	\path [<-] (B) edge (A);
	\end{tikzpicture}
	\end{array}}(t) \\ \nonumber 
	& = - 
  {\scriptsize 
	\begin{array}{c}
	\begin{tikzpicture}
	\node[cross out,draw=black] (x) at (0,0) {};
	\node[shape=circle,draw=black] (A) at (0,0) {};
	\node[shape=circle,draw=black] (B) at (.5,0) {};
	\node[shape=circle,draw=black,fill=blue!20] (C) at (1,-0.25) {};
	\node[shape=circle,draw=black,fill=blue!20] (D) at (1,0.25) {};
	\path [->] (A) edge node[left] {} (B);
	\path [->] (B) edge node[left] {} (C);
	\path [->] (B) edge node[left] {} (D);
	\end{tikzpicture}
	\end{array}}(t)  - 
		{\scriptsize 
	\begin{array}{c}
	\begin{tikzpicture}
	\node[cross out,draw=black] (x) at (0,0) {};
	\node[shape=circle,draw=black] (A) at (0,0) {};
	\node[shape=circle,draw=black] (B) at (.5,0) {};
	\node[shape=circle,draw=black] (C) at (1,0) {};
	\node[shape=circle,draw=black,fill=blue!20] (D) at (1.5,0) {};
	\path [->] (B) edge node[left] {} (C);
	\path [->] (C) edge node[left] {} (D);
	\path [<-] (B) edge (A);
	\end{tikzpicture}
	\end{array}}(t)
	+ {\scriptsize
	\begin{array}{c}
	\begin{tikzpicture}
	\node[cross out,draw=black] (x) at (0,0) {};
	\node[shape=circle,draw=black] (A) at (0,0) {};
	\node[shape=circle,draw=black] (B) at (.5,0.3) {};
	\node[shape=circle,draw=black,fill=blue!20] (C) at (.5,-.3) {};
	\node[shape=circle,draw=black,fill=blue!20] (D) at (1,0.3) {};
	\path [->] (A) edge node[left] {} (B);
	\path [->] (A) edge node[left] {} (C);
	\path [->] (B) edge (D);
	\end{tikzpicture}
	\end{array}}(t)
    \\ \nonumber & = 
    -2 {\scriptsize 
	\begin{array}{c}
	\begin{tikzpicture}
	\node[cross out,draw=black] (x) at (0,0) {};
	\node[shape=circle,draw=black] (A) at (0,0) {};
	\node[shape=circle,draw=black] (B) at (.5,0) {};
	\node[shape=circle,draw=black,fill=blue!20] (C) at (1,0) {};
	\path [->] (A) edge node[left] {} (B);
	\path [->] (B) edge node[left] {} (C);
	\end{tikzpicture}
	\end{array}}(t) +  \left( {\scriptsize 
	\begin{array}{c}
	\begin{tikzpicture}
	\node[cross out,draw=black] (x) at (0,0) {};
	\node[shape=circle,draw=black] (A) at (0,0) {};
	\node[shape=circle,draw=black,fill=blue!20] (B) at (.5,0) {};
	\path [->] (A) edge node[left] {} (B);
	\end{tikzpicture}
	\end{array}}(t) \right)^2 - {\scriptsize 
	\begin{array}{c}
	\begin{tikzpicture}
	\node[cross out,draw=black] (x) at (0,0) {};
	\node[shape=circle,draw=black] (A) at (0,0) {};
	\node[shape=circle,draw=black] (B) at (.5,0) {};
	\node[shape=circle,draw=black] (C) at (1,0) {};
	\node[shape=circle,draw=black,fill=blue!20] (D) at (1.5,0) {};
	\path [->] (B) edge node[left] {} (C);
	\path [->] (C) edge node[left] {} (D);
	\path [<-] (B) edge (A);
	\end{tikzpicture}
	\end{array}}(t) \nonumber \\ 
	& \quad + \left(
	{\scriptsize \begin{array}{c}
	\begin{tikzpicture}
	\node[cross out,draw=black] (x) at (0,0) {};
	\node[shape=circle,draw=black] (A) at (0,0) {};
	\node[shape=circle,draw=black,fill=blue!20] (B) at (.5,0) {};
	\path [->] (A) edge node[left] {} (B);
	\end{tikzpicture}
	\end{array}}(t)  \right) \left(
{\scriptsize 
	\begin{array}{c}
	\begin{tikzpicture}
	\node[cross out,draw=black] (x) at (0,0) {};
	\node[shape=circle,draw=black] (A) at (0,0) {};
	\node[shape=circle,draw=black] (B) at (.5,0) {};
	\node[shape=circle,draw=black,fill=blue!20] (C) at (1,0) {};
	\path [->] (A) edge node[left] {} (B);
	\path [->] (B) edge node[left] {} (C);
	\end{tikzpicture}
	\end{array}}(t) \right) \nonumber
\end{align}
Here the first equality is obtaining by separating the sink. The second line follows from \eqref{cuttingsources} and on the third line \eqref{secbranch} has been used, together with the observation \eqref{productrule} that branches emanating from a single source factorize into the product of the branches.

\section{Examples: sample networks}

\begin{figure}
	\includegraphics[width=\columnwidth]{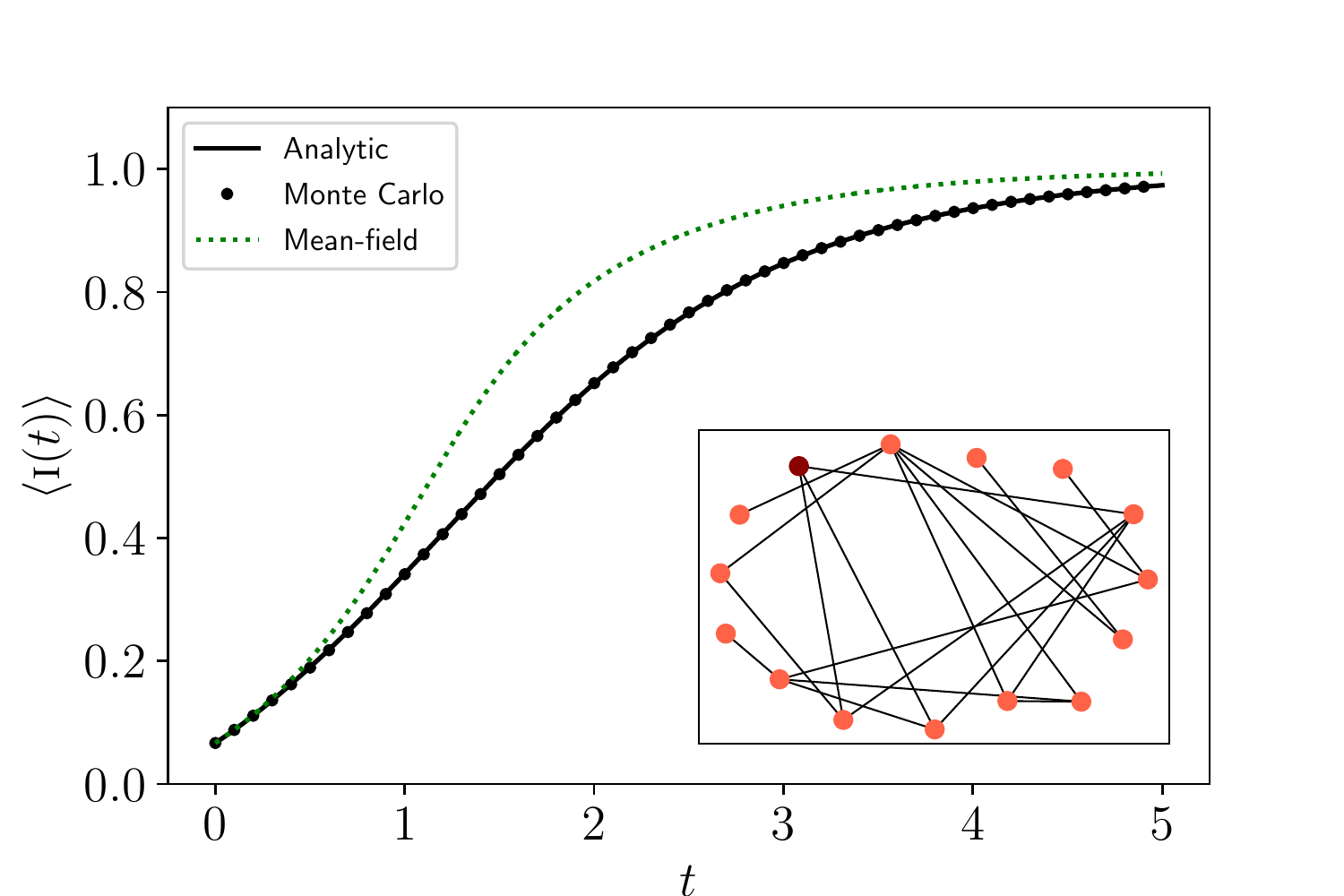} \\
    \includegraphics[width=\columnwidth]{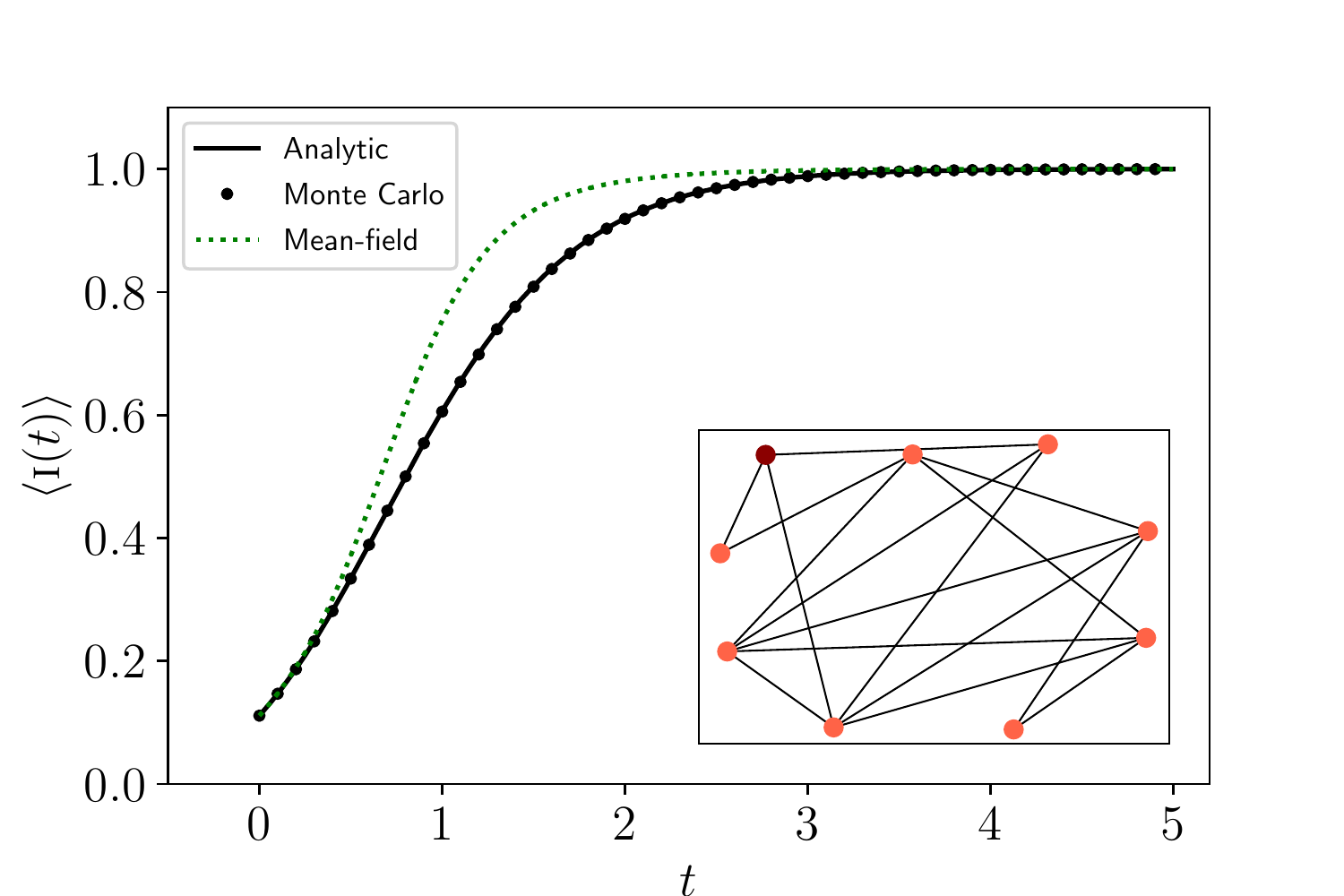} \\
	\includegraphics[width=\columnwidth]{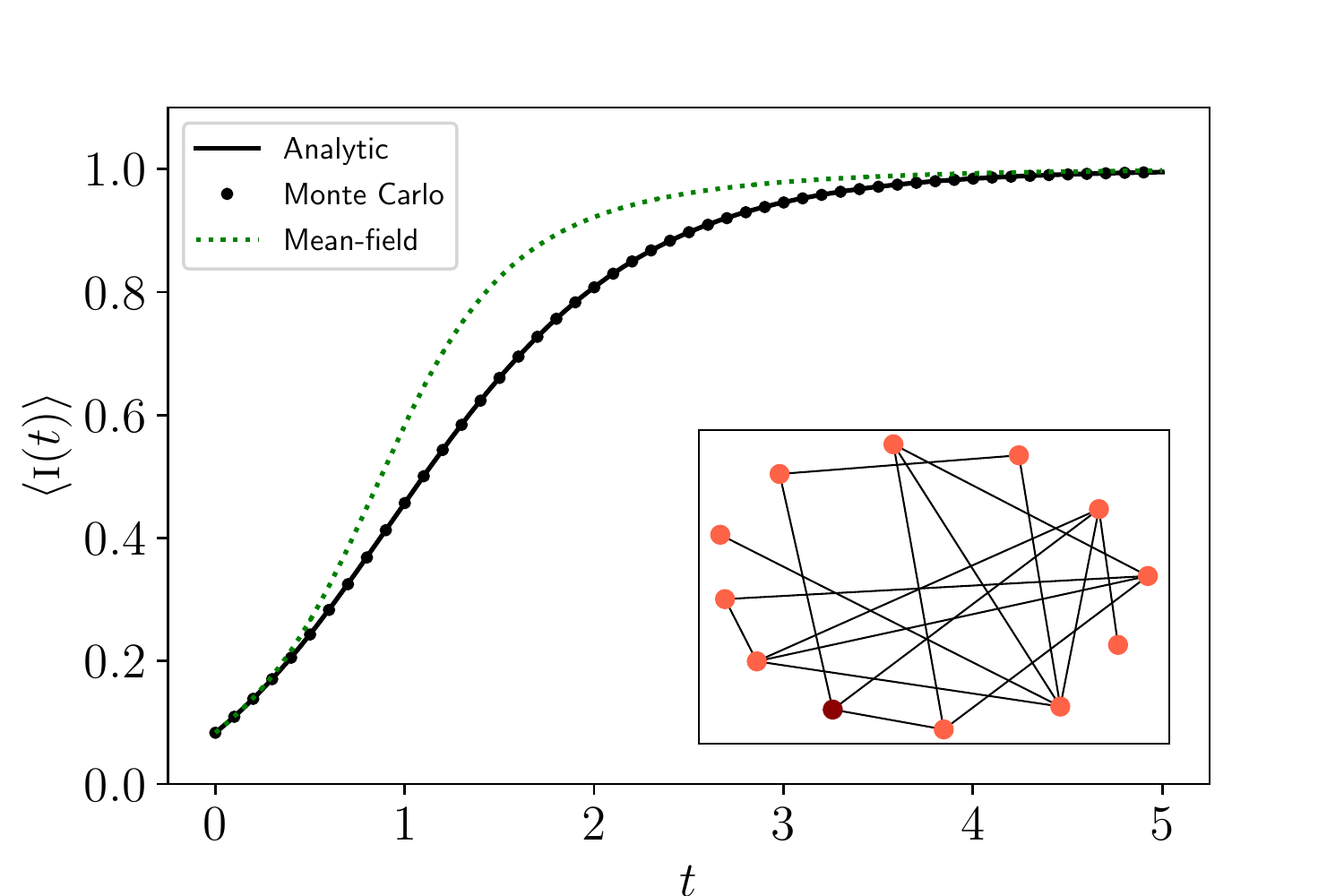}
	\caption{\label{fig:FF} The prevalence $\vev{\I(t)}$ for three selected networks: the graph of Florentine Family relations \cite{breiger1986cumulated} with 15 nodes and 20 edges (top), a randomly generated Newman-Watts-Strogatz small-world network \cite{newman1999renormalization,watts1998collective} with 9 nodes and 16 edges (center) and a random Erd\"os-R\'enyi graph \cite{erdos1960evolution} of 12 nodes and 17 edges (bottom). In each graph, the dark red node is the initial infected source. The solid black line gives the analytic result from summing all contributing diagrams, and agrees well with the numerical simulations (black dots, averaged over $60\,000$ runs). The green dashed line shows the individual based mean-field approximation and is in poor agreement with the exact result.}
\end{figure}

Due to the large number of subgraphs contributing to \eqref{diagramsexp}, explicit computation becomes increasingly prohibitive. For this reason, we have created a Python package \footnote{Merbis, Wout; Lodato, Ivano (2021): Exact solutions of the SI model on networks. figshare. Software. \url{https://doi.org/10.6084/m9.figshare.14872182.v3} } that, given an input network with specified sources and sinks, computes all contributing subgraphs, their parental relations \eqref{parentdef} and the integrals \eqref{diagramintegral} analytically.
Here we present explicitly the exact solutions for the prevalence \eqref{prevalence} on three small sample networks in figure \ref{fig:FF}. We suppose that the node marked dark red is a source and compute the expectation values \eqref{expvalue} for all of the other nodes. 
Our algorithm, as detailed in appendix \ref{app:algorithm}, computes for each graph all contributing diagrams, finds their parent contributions and constructs the graph of parental relationships. This allows us to integrate \eqref{diagramintegral} starting from the smallest subgraph. The normalized sum over all diagrams with a single sink node gives the prevalence, which we plot as solid black lines in figure \ref{fig:FF}. 

\begin{figure*}
    \includegraphics[width=.68\columnwidth]{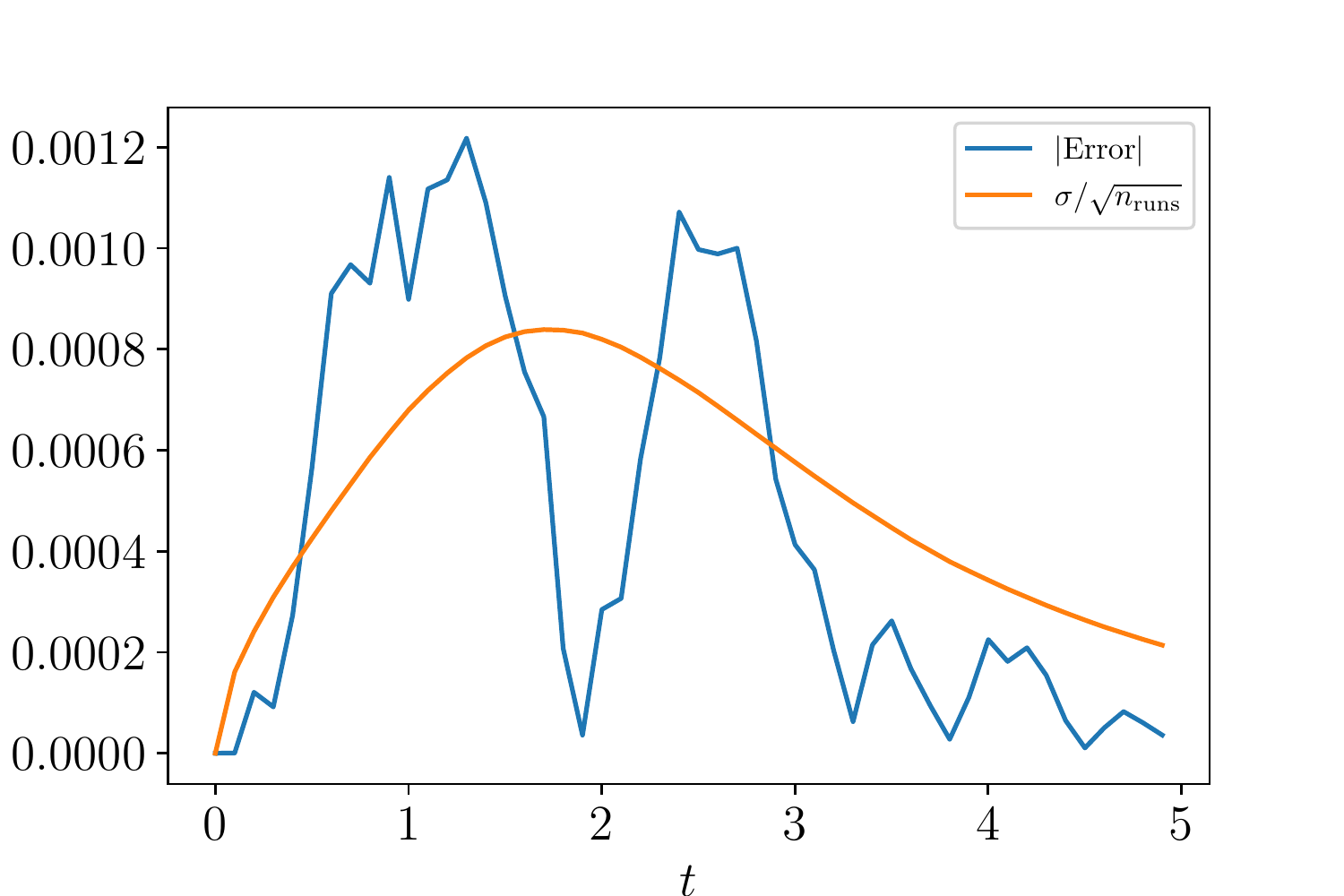}
	\includegraphics[width=.68\columnwidth]{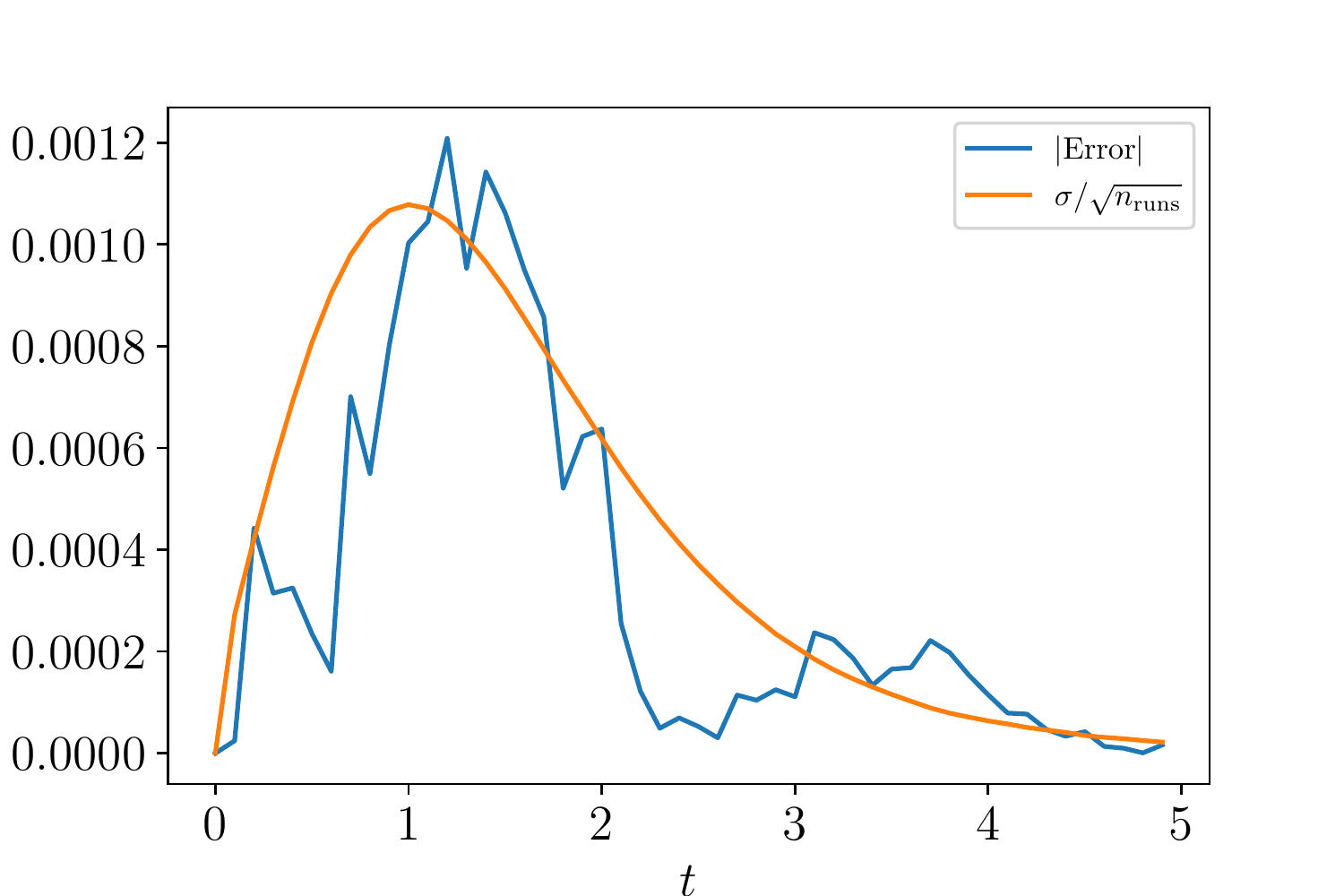}
	\includegraphics[width=.68\columnwidth]{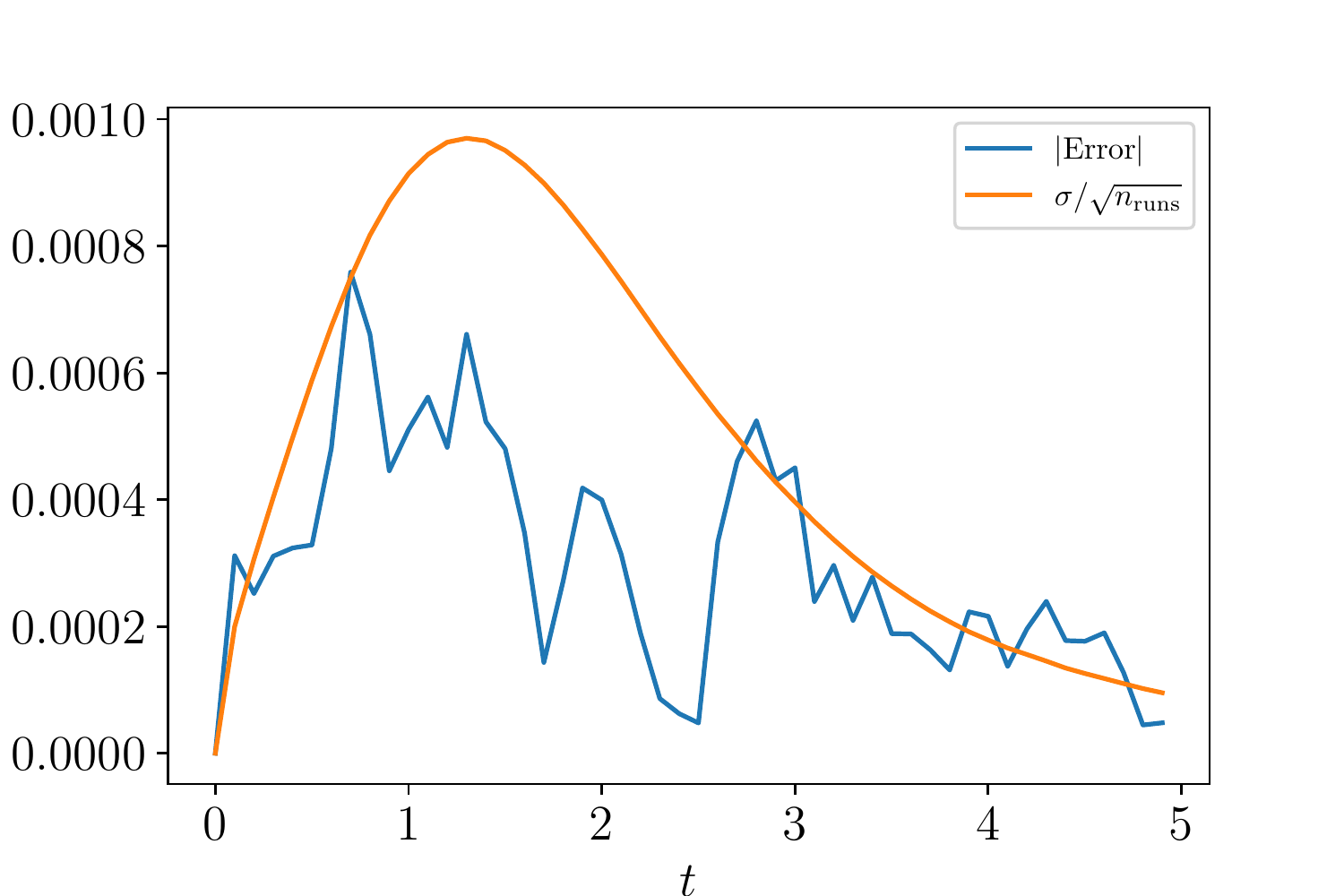}
	\caption{\label{fig:error} The absolute value of the difference between the simulated averages and the analytical function ($|{\rm Error}|$) compared to $\sigma /\sqrt{n}$, where $n$ is the number of simulation runs and $\sigma$ the standard deviation at each moment in time. The networks considered are Florentine Families (left), random small-world (center), random Erd\"os-R\'enyi (right).}
\end{figure*}

Our analytical results (black line) are in perfect agreement with Monte-Carlo simulations (black dots). These were performed by initializing the networks in the same initial state as in the analytic computation. Then, the Markovian dynamics is simulated by taking small time-steps where susceptible nodes connected to infected nodes are infected with a small infection probability $\tau$. As this is the only parameter in the model, $\tau$ effectively plays the role of the time-step $\Delta t$ for a dimensionless time. We choose the value of $\tau=0.002$, small enough to guarantee that at most a single node will flip its state in any one time-step. 
We simulate $n = 60\,000$ spreading trajectories over the same network and record the infection averages. The difference between the simulated average and the analytical formula is of the order of $\sigma /\sqrt{n}$ with $\sigma $ the standard deviation of the simulated runs, as shown in figure \ref{fig:error}. This is a good indication that the averages converge to the exact result in the limit of infinite simulation runs.
Note also that analytic and Monte-Carlo results clearly depart from individual-based mean field approximation (green dashed line), which gives an overestimation of the prevalence for finite $t$, as can be seen in figure \ref{fig:FF}.

\begin{figure*}
	\includegraphics[width=.68\columnwidth]{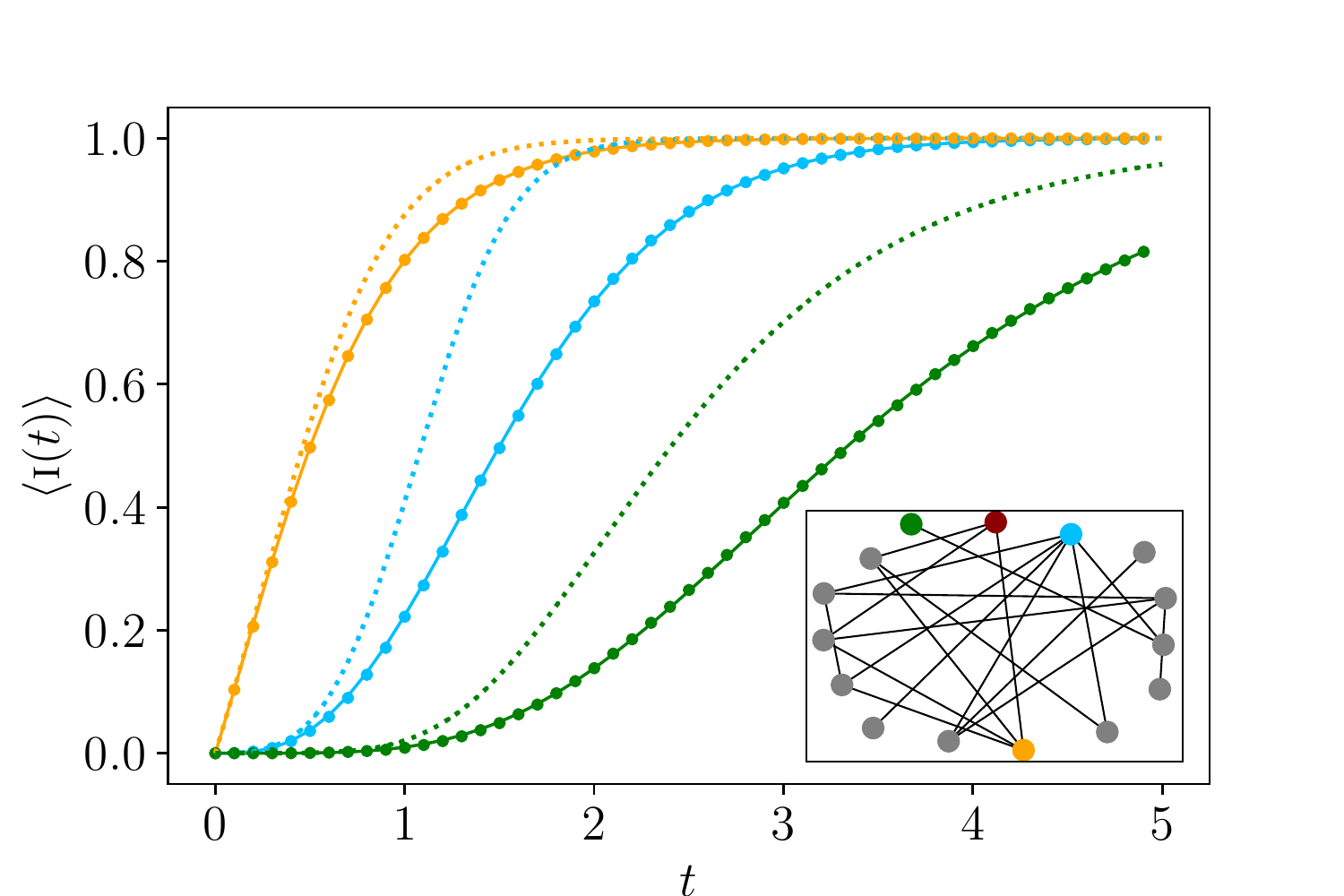}
	\includegraphics[width=.68\columnwidth]{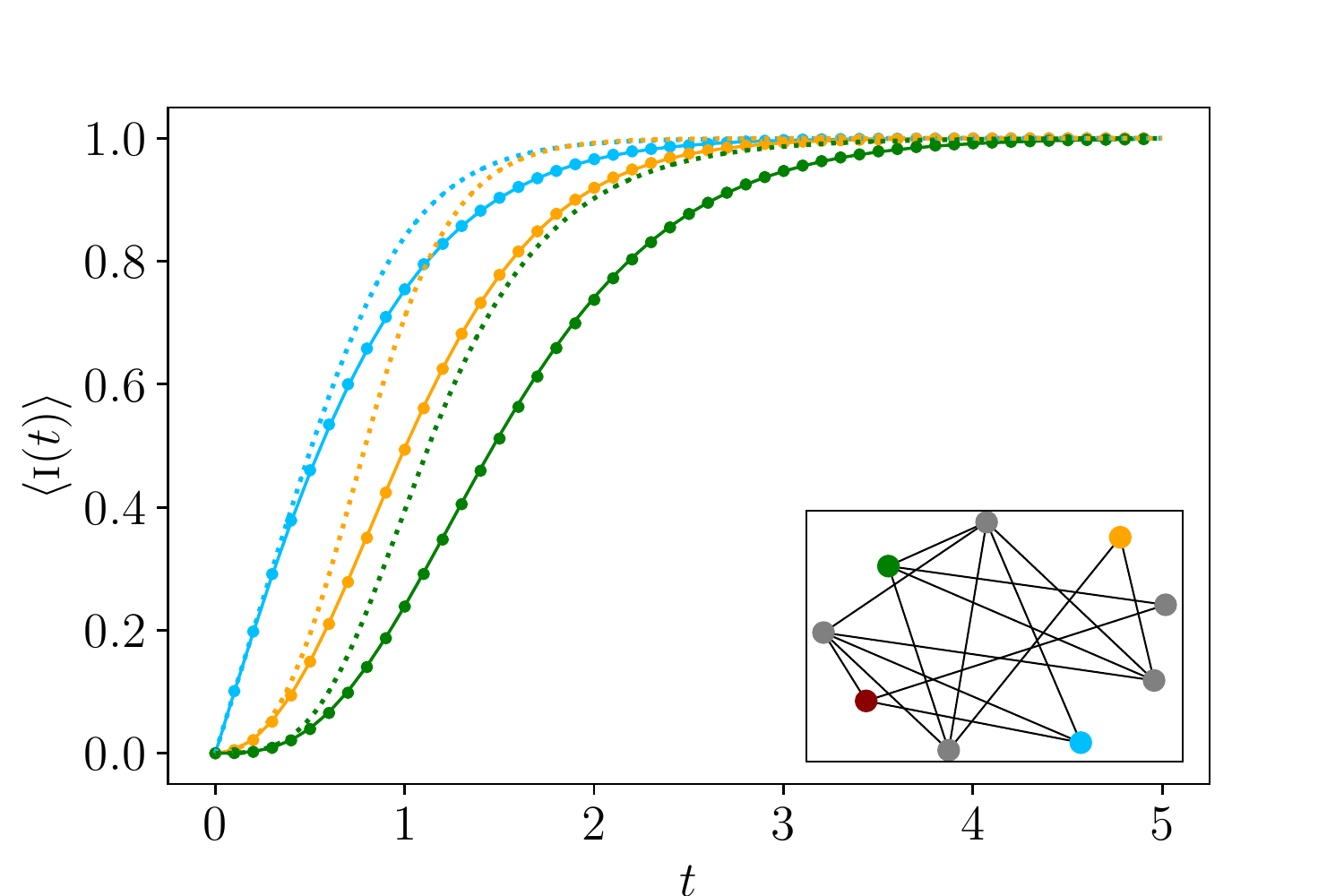}
	\includegraphics[width=.68\columnwidth]{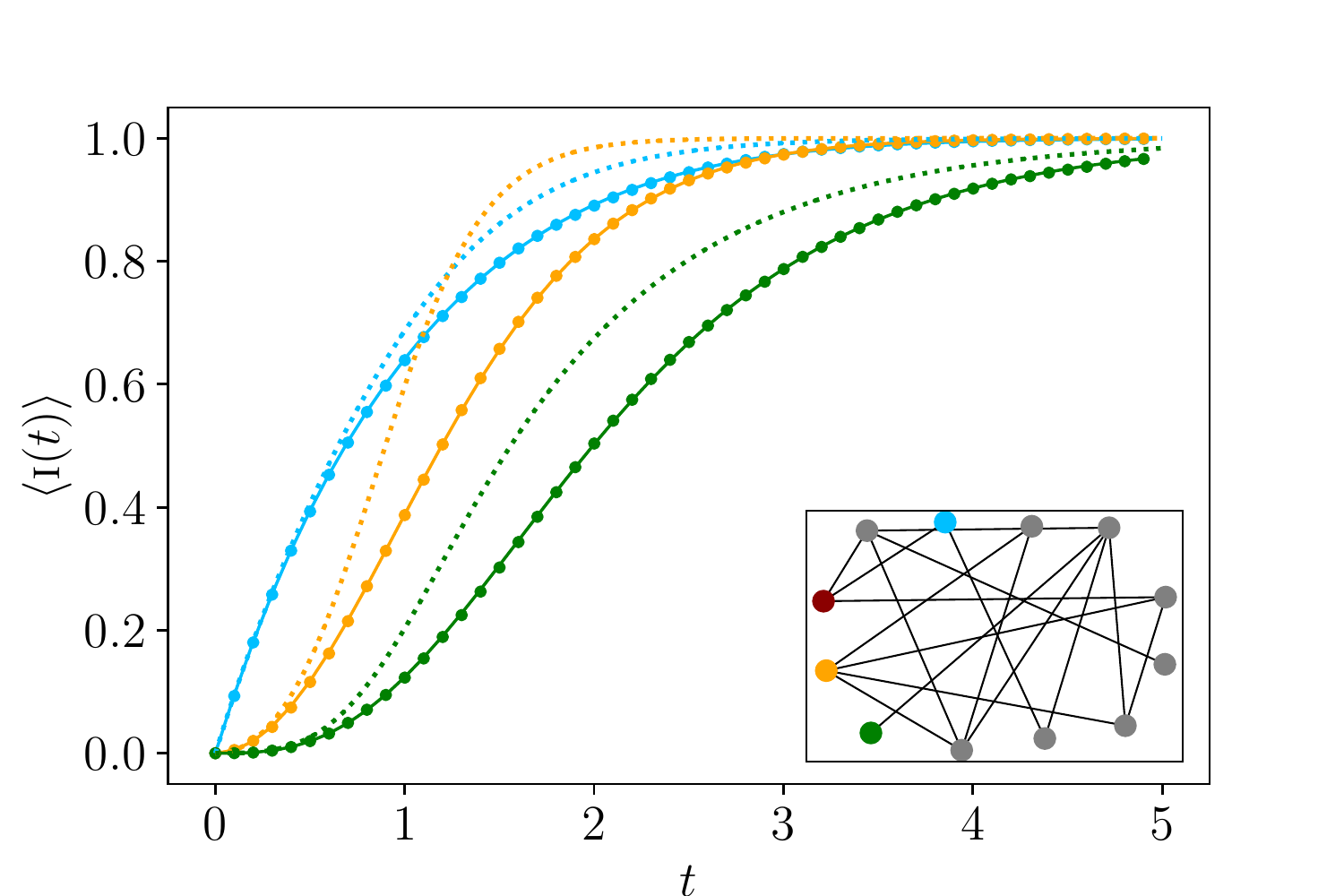}
	\caption{\label{fig:indFF} Expectation values for selected nodes from the Florentine Families graph (left), a random small world network (center) and a random Erd\"os-R\'enyi graph (right). The dark red node denotes the source node, and the blue, orange and green curves correspond to the like-colored nodes. The solid lines always correspond to the analytical expression obtained by summing all contributing diagrams. The dots correspond to the average of $60\,000$ Monte Carlo simulation runs. The dotted lines are the mean-field predictions, obtained by numerically integrating \eqref{meanfield}, and give overestimates of the spreading process.}
\end{figure*}
Besides the average expectation value for the full graph, we have also obtained the expectation value for individual nodes in the network. In figure \ref{fig:indFF} we plot the analytical result (solid line), the Monte-Carlo averages (dots) and the individual based mean-field approximations (dashed line) for three selected nodes in each network. The mean-field approximations are obtained by standard methods (see for instance \cite{newman2018networks}). The nodes in the network are assumed to be statistically independent, such that $\vev{\I^i \S^j} = \vev{\I^i} \vev{\S^j}$ at each moment in time. The individual-based mean-field approximation then consists out of the coupled set of ordinary differential equations (ODEs):
\begin{equation}\label{meanfield}
    \partial_t n_\I^i(t) = \sum_{j=1}^{N} A^{ij} (1- n_\I^i(t)) n_\I^j(t)\,.
\end{equation}
Here $n_\I^i(t)$ is the (mean-field) probability of node $i$ to be infected at time $t$. This set of equations is easily integrated numerically using a standard ODE integration routine (\texttt{odeint}). The approximation can by improved by including the dynamics for pairs and triples, closing the equations at higher orders \cite{kiss2017mathematics}, though our exact approach captures all correlations present in the network.

\section{Conclusion}
In this paper, we have presented a general formalism which allows for obtaining analytic solutions for the expectation values, as well as higher momenta, for the $\SI$ spreading process on general network. We have decomposed the time-dependent probability vector into a sum over subgraph diagrams, each of which can be computed systematically. We uncovered a set of symmetry relations, which relate contributions from diagrams among each other. Our methods are implemented numerically in \cite{Note2} and give results which match with Monte-Carlo simulations for several small sample networks of differing topologies. 

Even for the relatively small networks analyzed in the previous section, the explicit computation still involves performing thousands of integrals. It is a difficult task to determine exactly how the number of contributing diagrams scales with system size, as this depends strongly on the network topology. In general, the addition of a new edge could potentially double the number of contributing subgraphs, which will then scale exponentially with the number of edges.
Using our formalism, one can track only the necessary diagrams contributing to the spreading and systematically organize their contributions by number of edges. In fact, since any diagram with $m$ edges will start contributing at the order $t^m$ in the expansion \eqref{diagramsexp}, only the smallest contributing subgraphs need be taken into account for an early time estimate of the spreading process. Furthermore, the uncovered symmetry relations and combinations thereof further reduce the number of independent diagrams to be computed.
For example, for the three sample networks in figure \ref{fig:FF}, the number of inequivalent diagrams, including all parents are 9314 (top), 21\,376 (center) and 6055 (bottom), out of which only 2274, 1082 and 358 can \emph{not} be reduced to smaller diagrams using the symmetry relations presented. More symmetry relations may exist, which could further reduce the number of independent diagrams needed to compute the expectations values analytically. For instance, it would be interesting to study relations among contributions of diagrams with double directed edges.  

The formalism presented in this letter can be applied to different compartmental models of epidemiology and other out-of-equilibrium stochastic systems on networks \cite{boccaletti2006complex,gleeson2013binary,PastorSatorras2015}. It would be interesting to determine under which conditions exact solutions of such models can be found. We suspect this to be the case for any system where the graph of diagrams is acyclic, such that any diagrams contribution can be traced back to the smallest diagram of a single source.

\appendix

\section{Infinite time limit}\label{app:latetime}

We will now prove that the contribution $a_G(t)$ of each diagram $G$ reaches asymptotically a constant integer value $k_G \neq 0$. 
As a first step, we explicitly check that the function of $t$ for the unique network with $m=1$ edge converges to a constant (integer) value. This is trivial since the expectation value for a one-edge network takes on the form $1-e^{-t}$ which converges to $1$ for $t\to\infty$.
Next we use the inductive hypothesis that all diagrams $G_{m-1}$ with $m-1$ edges converge to a non-zero constant for $t\to\infty$, i.e:
\begin{equation}\label{indhyp}
    a_{G_{m-1}}(t) = k_{G_{m-1}} + f_{G_{m-1}}(t)\,, 
\end{equation}
such that $\lim_{t \to +\infty} f_{G_{m-1}}(t)=0 $ and $k_{G_{m-1}} \neq 0$.
Now we want to show, using the above hypothesis, that all diagrams with $m$ edges converge to a non-zero constant asymptotically. Let us consider the integral
\begin{align}
    a_{G_m}(t) & = e^{-ct}\int_0^t e^{c\,s} p_{G_{m}}(s) \; {\rm d} s  \\ \nonumber
    & = e^{-ct}\int_0^t e^{c\,s} \sum_{K \in P(G_m)} \alpha_{G K} a_{K}(s) \; {\rm d} s 
\end{align}
where $P(G_m)$ are all parents of the $m$-edged diagram $G_m$, which necessarily have $m-1$ edges. Plugging in the inductive hypothesis \eqref{indhyp}, we get:
\begin{equation}
    a_{G_m}(t)=e^{-ct}\int_0^t e^{c\, s} \sum_{K \in P(G_m)} \alpha_{G K} \big(\, k_K +f_K(s)\big) \, {\rm d} s
\end{equation}
where $\alpha_{G K}$ is the multiplicity and relative sign of the $(m-1)$-edged parent diagram $K$ and $k_K$ denotes the limiting value of that parent. Now we can separate two contributions:
\begin{align}
    a_{G_m}(t) = & \, \sum_{K \in P(G_m)} \alpha_{GK} \frac{ k_K}{c} \big(1-e^{-ct}\big) \\ \nonumber
    & + \sum_{K \in P(G_m)} \alpha_{GK} e^{-ct}\int_0^t e^{c\, s} f_K(s) \; {\rm d}s
\end{align}
Note that the first term contains a finite contribution when $t\to\infty$. The second term converges to 0 for $t\to\infty$. To see this, consider:
\begin{align}
    & e^{-ct}\int_0^t e^{c\,s} f(s) \; {\rm d} s =e^{-ct}\big(F(t)-F(0)) \\ \nonumber
    & \text{with: } F^\prime(t)=e^{ct}f(t)
\end{align}
The asymptotic limit of $e^{-ct} F(t)$ is evaluated as:
\begin{align}
    \lim_{t\to\infty} e^{-ct} F(t)&=\lim_{t\to\infty} \frac{F(t)}{e^{ct}}=\lim_{t\to\infty} \frac{F^\prime(t)}{ce^{ct}}
    \nonumber\\
    &=\lim_{t\to\infty}\frac{e^{ct}f(t)}{ce^{ct}}=\lim_{t\to\infty}\frac{f(t)}{c}=0
\end{align}
where to obtain the second equality we used L'H\^opital's rule and in the second line the inductive assumption \eqref{indhyp} was used. In addition, $\lim_{t\to\infty} e^{-ct} F(0) =0$, since $F(0)$ can only be a constant. Hence we have shown that all diagrams, regardless of the number of edges, will converge asymptotically to a non-zero constant $k_G$, such that
\begin{equation}\label{limitvalue}
    k_G = \lim_{t\to\infty} a_G(t) = \frac{1}{c} \sum_{K \in P(G)} \alpha_{GK} k_K\,. 
\end{equation}
Next, we prove that $k_G$ is a non-zero integer, i.e. $k_G \in \mathbb{Z}$. First, the one-edge diagram $G_1$ has asymptotic value $k_{G_1}=1$. Let us assume (inductive hypothesis) that all diagrams with $m-1$ single-directed edges have limit $\pm 1$. If no double-directed edges are present the number of parents will always be equal to the number of edges going into the blue nodes in the original diagram, $c$. To see this, consider one such diagrams $G_{m}$ with $m$ edges: one can unequivocally obtain its parents $G_{m-1} \in P(G_m)$ by systematically removing one of the edges connecting to one of the sinks. As there are $c$ such edges, there will be $c$ parents.  Finally, the sign of each parent contribution to \eqref{limitvalue} is equal to the sign of $G$, ${\rm sign}(\alpha_{GK}k_K) = {\rm sign}(G)$. This is due to the fact that parents $K$ with signs opposing $G$ will transition into $G$ by one of the rules which induce a sign change. This sign is then contained in $\alpha_{GK}$ and the resulting product $\alpha_{GK} k_K$ will have the same sign as $G$. By induction, then, for the $m$-edged diagram $G$, we have that $\sum_{K\in P(G)} \alpha_{GK} k_K = {\rm sign(G)}c$ and hence \eqref{limitvalue} gives:
\begin{equation}\label{kG}
    k_G = {\rm sign}(G) \;.
\end{equation}
If double directed edges exist in a diagram $G_D$ with $c$ edges to blue nodes, then there could be more than $c$ parents. The reason for this is that each direction on the double directed edge will correspond to different parent diagrams. We can then write down all possible diagrams with the same topology as $G_D$, but where the $D$ double directed edges take on a given orientation, consistent with the graphs source and sink configuration. The number $n_o$ of these diagrams is bounded $n_o \le 2^D$, since certain double-arrows may exist only conditionally to the existence of other double arrows. Now, each diagram among these $n_o$ will have no double-directed edges and $c$ edges to blue nodes, and hence, $c$ parents. Each of these parents will have limit ${\rm sign}(G_D)$ by the above argument, so finally:
\begin{equation}
    k_{G_D}= {\rm sign}(G_D) \, n_o\,.
\end{equation}
Hence any diagram will, regardless of its topology, converge to a non-zero integer. This integer corresponds to the number of possible orientations of the double directed edges in the diagram (or equal 1 if there are no double directed edges). For example, the first diagram with a double directed edge \eqref{excitation} converges to $-2$, since its sign \eqref{sign} is negative and there are two possible orientations of the double directed edge (up or down). Each choice of orientation has 2 parents with limit $-1$, resulting in a total of 4 parents. The diagram has $c=2$, such that \eqref{limitvalue} gives $-2$.

\section{Proof of \eqref{cuttingsources}}\label{app:cuttingproof}

\begin{figure*}[t]
\begin{minipage}{\textwidth}
\begin{algorithm}[H]
	\caption{Compute expectation value}
	\label{ComputeExpVal}
	\begin{algorithmic}[1]
	\Procedure{ComputeExpVal}{$G,s,i$}
	\State {\bf Input:} graph $G$, source node $s$, sink node $i$
	\State {\bf Output:} Expectation value $\vev{\I^i(t)}$
	\State $\textit{contributions} \gets \Call{ContributingDiagrams}{G,s,i}$
	\State $\textit{GG.nodes} \gets$ inequivalent diagrams from $\textit{contributions}$
	\ForAll{ $\textit{diagram} \in \textit{GG.nodes}$} \Comment{Run from largest to smallest diagram}
	\State $\textit{parents} \gets \Call{GetParents}{\textit{diagram}}$
	\State Append \textit{parents} to \textit{GG.nodes}
	\ForAll{$p \in \textit{parents}$} 
	\State add edge $(\textit{diagram},p)$ to \textit{GG.edges} with weight $w = \Call{Sign}{\textit{diagram}} *\Call{Sign}{\textit{p}}$
	\EndFor
	\EndFor
	\State $a(t) \gets 1$ for the initial configuration $s$
	\ForAll{$\textit{diagram} \in \textit{GG.nodes}$} \Comment{Run from smallest to largest diagram}
	\State $p(t) \gets  $ sum over $w*a(t)$ for all neighbors \textit{p} of \textit{diagram}
	\State $c \gets $ total degree of blue nodes in \textit{diagram}
	\State $a(t) \gets e^{-c t}\int_0^t e^{c \,s} p(s) \textrm{d}s$
	\EndFor
	\State $\textit{result} \gets $ sum over $a(t)$ for each $\textit{diagram} \in \textit{contributions}$
	\State \Return \textit{result}
	\EndProcedure
	\end{algorithmic}
\end{algorithm}
\end{minipage}
\end{figure*}

Here we will provide a proof of the `cutting off sources' symmetry relation, presented in section \ref{sec:cuttingsources}. For notational convenience, we will use the functions $a_{G^{+}}(t), p_{G^+}(t)$ and $a_G(t)$ to denote the diagrams contributions. Here $G^+$ is the graph obtained from $G$ by extending the source by one edge. Equation \eqref{cuttingsources} of the main text then reads:
\begin{equation}\label{genrule2}
    (c-1)a_{G^+}(t)=
	p_{G^+}(t)
	-
	a_G(t) \,.
\end{equation}
Let us first start by considering $a_{G^+}(t)$, recalling the definition of the contribution as
\begin{equation}\label{genrule2bis}
a_{G^{+}}(t) =\sum_{n=m+1}^{+\infty} a_{G^{+}}[n]\frac{t^n}{n!}\,.
\end{equation}
Here $m$ is the number of edges of the diagram $G$.
The coefficients $a_{G^{+}}[n]$ count the number of ways the graph $G^{+}$ can be constructed by applying $\Q$ $n$ times on the initial configuration. The difference between the diagrams $G^{+}$ and $G$ is the presence of a chain of length one emanating from the source. Hence to construct the diagram $G^+$, first the chain must be made, and then the diagram $G$ can be created.
So for the operator $\Q$ to generate $G^{+}$ in $n>m$ steps, it must be applied \emph{at least} once to create the chain, and \emph{at most} $n-1$ times to create the $m$ edges of $G$.  The total number of ways to construct $G^+$ is then given by the sum over all ways to construct the chain in $1\leq i \leq n-m$ steps times the number of ways to construct $G$ in $n-i$ steps, such that:
\begin{equation}\label{aGplus}
    a_{G^{+}}[n] = \sum_{i=1}^{n-m} 
    {\scriptsize 
	\begin{array}{c}
	\begin{tikzpicture}
	\node[cross out,draw=black] (x) at (0,0) {};
	\node[shape=circle,draw=black] (A) at (0,0) {};
	\node[shape=circle,draw=black,fill=blue!20] (B) at (0.5,0) {};
	\path [->] (A) edge node[left] {} (B);
	\end{tikzpicture}
	\end{array}}[i] \cdot a_G[n-i] \;.
\end{equation}
Now, the chain of length one $G_1$ satisfies the recursion relation 
\begin{equation}
    {\scriptsize 
	\begin{array}{c}
	\begin{tikzpicture}
	\node[cross out,draw=black] (x) at (0,0) {};
	\node[shape=circle,draw=black] (A) at (0,0) {};
	\node[shape=circle,draw=black,fill=blue!20] (B) at (0.5,0) {};
	\path [->] (A) edge node[left] {} (B);
	\end{tikzpicture}
	\end{array}}[n] = - {\scriptsize 
	\begin{array}{c}
	\begin{tikzpicture}
	\node[cross out,draw=black] (x) at (0,0) {};
	\node[shape=circle,draw=black] (A) at (0,0) {};
	\node[shape=circle,draw=black,fill=blue!20] (B) at (0.5,0) {};
	\path [->] (A) edge node[left] {} (B);
	\end{tikzpicture}
	\end{array}}[n-1]\,,
\end{equation}
with the boundary conditions that $a_{G_1}[n=1]=1$. This implies that  $a_{G_1}[n]=(-1)^{n+1}$ so that equation \eqref{aGplus} becomes 
\begin{equation}
\label{genrule2tris}
    a_{G^{+}}[n]=\sum_{i=1}^{n-m} (-1)^{i+1} \, a_{G}[n-i]\,.
\end{equation}
Consider now the sum:
\begin{align}
    & \quad a_{G^{+}}[n]+a_{G^{+}}[n-1] \\
    &=\sum_{i=1}^{n-m}(-1)^{i+1}\cdot a_{G}[n-i] +\sum_{i=1}^{n-m-1}(-1)^{i+1}\cdot a_{G}[n-i-1]
    \nonumber\\
    &=\sum_{i=1}^{n-m}(-1)^{i+1}\cdot a_{G}[n-i] +\sum_{i=2}^{n-m}(-1)^{i}\cdot a_{G}[n-i]
    \nonumber\\
    &=a_G[n-1] \nonumber
\end{align}
or equivalently:
\begin{equation}
    \label{genrule2fin}
    a_{G^{+}}[n+1]=a_G[n] - a_{G^{+}}[n]
\end{equation}
Next, we multiply this equation by $t^n/n!$ and sum over $n$. The right hand side immediately turns into the corresponding contributions as a function of $t$. The left hand side can instead be expressed as the time derivative of $a_{G^+}(t)$:
\begin{align}
    & \qquad a_G(t)- a_{G^{+}}(t)=\sum_{n=m}^{+\infty} a_{G^{+}}[n+1]\frac{t^n}{n!}  \nonumber \\ 
    & =\partial_t \Big(\sum_{n=m}^{+\infty} a_{G^{+}}[n+1]\frac{t^{n+1}}{(n+1)!}\Big)=\partial_t a_{G^{+}}(t)
    \label{diffeqn} \\
    &= -c\, a_{G^{+}}(t)+  p_{G^{+}}(t) \,.
\end{align}
Here we have used the dynamical equation \eqref{dyneqn} in the main text to obtain the last equality. A simple rewriting now reproduces the equation \eqref{genrule2} for a generic graph $G$.
One can also obtain a relation between $a_{G^{+}}(t)$ and $a_{G}(t)$ independent of the parents $p_{G^+}(t)$ from solving the differential equation in the first line of \eqref{diffeqn} as
\begin{equation}
    a_{G^{+}}(t)=e^{-t}\int_0^t e^{s} a_G(s) \, {\rm d}s \,.
\end{equation}

\begin{figure*}[t]
\begin{minipage}{\textwidth}
\begin{algorithm}[H]
	\caption{Collect contributing diagrams} 
	\label{ContributingDiagrams}
	\begin{algorithmic}[1]
        \Procedure{ContributingDiagrams}{$G,s,i$}
        \State $\textbf{Input:} \textrm{ graph } G, \textrm{ source node } s, \textrm{ sink node } i$
        \State $\textbf{Output:}$ list of contributing diagrams $\textit{contributions}$
        \State $\textit{edges} \gets (u,v)$ such that $ \exists \textit{ path}(s,i) \in G$ passing through $(u,v)$
        \State $\textit{contributions} \gets $ Digraph $\gets \textit{edges}$
        \For{$\textit{diagram} \in \textit{contributions}$}
        \For{$\textit{edge} \in \textit{diagram}$}
        \State remove $\textit{edge}$
        \If {$\textit{diagram} =$ connected} 
        \ForAll{$j \in \textit{diagram}$}
        \If {$\nexists \textit{ path}(s,i)$ through $j$}
        \State remove node $j$ \Comment{Removes nodes not in the in-component of $i$}
        \EndIf
        \EndFor
        \State add $\textit{diagram}$ to $\textit{contributions}$ \Comment{Only add each unique configuration once}
        \EndIf
        \EndFor
        \EndFor
        \State \Return \textit{contributions}
        \EndProcedure
	\end{algorithmic} 
\end{algorithm}
\end{minipage}
\end{figure*}

\section{Analytical methods and code} 
\label{app:algorithm}

\begin{figure*}[t]
\begin{minipage}{\textwidth}
\begin{algorithm}[H]
	\caption{Find the parent diagrams} 
	\label{GetParents}
	\begin{algorithmic}[1]
        \Procedure{GetParents}{$\textit{diagram}$}
        \State $\textbf{Input:} \textrm{ diagram } $
        \State $\textbf{Output:}$ list of parent diagrams $\textit{parents}$
        \For{$\textit{b} \in \textit{diagram}$ such that $\textit{state}[b] = $ blue}
        \State $\textit{neighbors} \gets$ list of neighbors of $b$
        \For{$i \in \textit{neighbors} $}
        \State remove edge $(i,b)$ \Comment{If $b$ is now disconnected, remove $b$}
        \If{$k_{\rm out}^i = 0$} \Comment{If there are no outgoing edges, the neighbor $i$ must be blue}
        \State $\textit{state}[i] = $ blue
        \State append $\textit{diagram}$ to $\textit{parents}$
        \EndIf
        \If{$n_{\rm out}^i \in n_{\rm in}^i$} \Comment{All outgoing edges are double directed: the neighbor can be both white and blue}
        \State fix directed edges of $\textit{diagram}$ and append to $\textit{parents}$ 
        \If{$\textit{state}[i] \neq $ source } \Comment{Sources can not change state to blue}
        \State $\textit{state}[i] = $ blue
        \State fix directed edges of $\textit{diagram}$ and append to $\textit{parents}$
        \EndIf
        \Else \Comment{There is at least one single directed outgoing edge and the neighbor must stay white}
        \State fix directed edges of $\textit{diagram}$ and append to $\textit{parents}$ 
        \EndIf
        \EndFor
        \EndFor
        \State \Return \textit{parents}
        \EndProcedure
	\end{algorithmic} 
\end{algorithm}
\end{minipage}
\end{figure*}

To compute the exact solution for the expectation value in a sample network we developed a numerical code in Python for the integration of each diagram, which combines the network functionality of \texttt{NetworkX} with the symbolic manipulation and integration functionality of \texttt{SymPy}. The algorithm to obtain the exact solution for $\vev{\I^i(t)}$ in \eqref{expvalue} consists out of three basic steps. First, we create a list of contributing diagrams as \texttt{NetworkX} DiGraphs by collecting all (labelled) subgraph configurations where the node $i$ is blue. Secondly, the graph of diagrams $GG$ for each inequivalent diagram in the list is constructed by finding the parents of all diagrams, starting from the largest subgraph down to the initial conditions. Grandparents are added iteratively and each diagram is stored as a node in $GG$. The multiplicities $\alpha_{GH}$ of topologically inequivalent parents are stored as the edge weights of $GG$.
Finally, $a_G(t)$ for each diagram in $GG$ is obtained by integrating eqn. \eqref{diagramintegral}. This is done starting from the smallest diagram up to the largest contributing diagram. The result is then the sum over all diagrams contributions obtained in the first step.

The pseudo-code for the algorithm to compute the expectation value is given in Algorithm \ref{ComputeExpVal}. It relies on a number of functions which we will expand upon below. $\textsc{ContributingDiagrams}(G,s,i)$ computes a list of all labelled subgraph configurations which contribute to the expectation value for the node $i$ to be infected at time $t$, given that node $s$ was initially infected. $\textsc{GetParents}(\textit{diagram})$ gives a list of parent diagrams which transition into $\textit{diagram}$ by applying the dynamical rules of the $\SI$ system. $\textsc{Sign}(\textit{diagram})$ simply computes the sign \eqref{sign} of a diagram and is needed if the dynamical rules generate a sign difference between the parent and the child diagram, for instance by closing a loop on a blue node.

\subsection{Contributing diagrams}

The list of contributing diagrams is constructed as follows. First, the graph $G$ is initialized by defining the states of the source node $s$ and the sink node $i$. Then all possible paths from $s$ to $i$ are stored in a list and a directed diagram is built from these paths, with only white nodes in between $s$ and $i$. The resulting diagram is an acyclic directed graph containing all possible paths from $s$ to $i$ corresponding to the largest possible diagram contributing to the expectation value. 
From this diagram, edges are removed systematically such that the resulting graph is connected and all remaining nodes stay within the in-component of the blue node $i$. If any node is not in the in-component of the node $i$, it is removed along with all edges connected to it. Each resulting connected diagram is added to the list of contributing diagrams. The pseudo-code for this algorithm is provided in Algorithm \ref{ContributingDiagrams}.

\subsection{The Get Parents function} 
\label{sec:get_parents}

In order to integrate any give diagram, one needs to recover the parents of that diagram. This can be done by tracing the rules of the $\SI$ model backwards. The parents of any given diagram necessarily have one edge less connected to a blue node. They can thus be obtained by systematically removing single edges connected to the sinks, while leaving all other edges intact. However, the neighbor connected to the removed edge might change state from white to blue, as the forward action of $\Q$ can project blue nodes into white ones. Special care must be made to include only actual parents in the list of parent diagrams, as some coloring of nodes can not be obtained from the forward action of $\Q$ on the initial conditions. For instance, it is impossible to have a diagram with two neighboring blue nodes (connected by an edge). Similarly, it is not possible to have a branch in the diagram without a blue node, as there would be no path from source to sink in that branch. 

To determine whether a candidate parent of a given diagram can actually exist, we exploit the directed edges representing the flow from sources to sinks. Specifically, we must check that for each neighbor $i$ of a sink, after removal of their edge, other outgoing edges from $i$ exist. If so, the node $i$ cannot be blue and must be white (sinks cannot have outgoing edges). If, on the other hand, the neighboring node has only incoming edges, it \textit{must} be turned blue, since the flow of information must end in a sink. Finally, there is a scenario where the node $i$ can be both blue \textit{or} white. This happens when the node has only double directed edges as outgoing edges, and so the set of out-neighbors of $i, n_{\rm out}^i$ is contained within the set of in-neighbors $n_{\rm in}^i$. In that case, both the diagrams with node $i$ white \textit{and} blue are valid parents. It could also be that some paths are no longer possible after removing the edge to a blue node, or after turning a white node blue. Therefore, the direction of the edges for the parent diagrams should be re-evaluated in each case. The pseudo-code for the algorithm which constructs all physical parents of any arbitrary diagram is given in Algorithm \ref{GetParents}.

Our code is freely available in the online public repository  \cite{Note3}, 
along with a list of contributions $a_G(t)$ for all $80\,332$ diagrams with up to 10 edges. The repository also includes several Jupyter notebooks where the example graphs of the main text are computed using our algorithms. These notebooks contain the exact analytical expressions for the infection prevalence plotted in Fig. \ref{fig:FF} of the main text.

\section{Explicit functions for the first few diagrams}
\label{app:diagrams}

For completeness we list here the explicit functions $a_G(t)$ corresponding to all diagrams with up to four edges. 

\begin{widetext}
\begin{align*}
    {\scriptsize \begin{array}{c}
	\begin{tikzpicture}
	\node[cross out,draw=black] (x) at (0,0) {};
	\node[shape=circle,draw=black] (A) at (0,0) {};	
	\node[shape=circle,draw=black,fill=blue!20] (B) at (.5,0) {};
	\path [->] (A) edge node[left] {} (B);
	\end{tikzpicture}
	\end{array}}(t) & = 1-e^{-t}\,, 
	\\
	{\scriptsize 
	\begin{array}{c}
	\begin{tikzpicture}
	\node[cross out,draw=black] (x) at (0,0) {};
	\node[shape=circle,draw=black] (A) at (0,0) {};
	\node[shape=circle,draw=black] (B) at (.5,0) {};
	\node[shape=circle,draw=black,fill=blue!20] (C) at (1,0) {};
	\path [->] (A) edge node[left] {} (B);
	\path [->] (B) edge node[left] {} (C);
	\end{tikzpicture}
	\end{array}}(t)  & = 1 - (1+ t)e^{-t} \,, 
    \\
	{\scriptsize 
	\begin{array}{c}
	\begin{tikzpicture}
	\node[cross out,draw=black] (x) at (0,0) {};
	\node[shape=circle,draw=black] (A) at (0,0) {};
	\node[shape=circle,draw=black,fill=blue!20] (B) at (.5,-.25) {};
	\node[shape=circle,draw=black,fill=blue!20] (C) at (.5,.25) {};
	\path [->] (A) edge node[left] {} (B);
	\path [->] (A) edge node[left] {} (C);
	\end{tikzpicture}
	\end{array}}(t) & = (1-e^{-t})^2 \,, 
\\
	{\scriptsize 
	\begin{array}{c}
	\begin{tikzpicture}
	\node[cross out,draw=black] (x) at (0,0) {};
	\node[shape=circle,draw=black] (A) at (0,0) {};
	\node[shape=circle,draw=black] (B) at (.5,0) {};
	\node[shape=circle,draw=black,fill=blue!20] (C) at (1,-0.25) {};
	\node[shape=circle,draw=black,fill=blue!20] (D) at (1,0.25) {};
	\path [->] (A) edge node[left] {} (B);
	\path [->] (B) edge node[left] {} (C);
	\path [->] (B) edge node[left] {} (D);
	\end{tikzpicture}
	\end{array}} (t) & =  1 - 2t e^{-t} - e^{-2t} \,, 
    \\
	{\scriptsize 
	\begin{array}{c}
	\begin{tikzpicture}
	\node[cross out,draw=black] (x) at (0,0) {};
	\node[shape=circle,draw=black] (A) at (0,0) {};
	\node[shape=circle,draw=black] (B) at (.5,0) {};
	\node[shape=circle,draw=black] (C) at (1,0) {};
	\node[shape=circle,draw=black,fill=blue!20] (D) at (1.5,0) {};
	\path [->] (A) edge node[left] {} (B);
	\path [->] (B) edge node[left] {} (C);
	\path [->] (C) edge (D);
	\end{tikzpicture}
	\end{array}}(t) & = 1 - (1+ t + \tfrac12 t^2)e^{-t} \,, 
\end{align*}
\begin{align*}
	{\scriptsize
	\begin{array}{c}
	\begin{tikzpicture}
	\node[cross out,draw=black] (x) at (0,0) {};
	\node[shape=circle,draw=black] (A) at (0,0) {};
	\node[shape=circle,draw=black] (B) at (.5,0.3) {};
	\node[shape=circle,draw=black,fill=blue!20] (C) at (.5,-.3) {};
	\node[shape=circle,draw=black,fill=blue!20] (D) at (1,0.3) {};
	\path [->] (A) edge node[left] {} (B);
	\path [->] (A) edge node[left] {} (C);
	\path [->] (B) edge (D);
	\end{tikzpicture}
	\end{array}} (t) = - {\scriptsize
	\begin{array}{c}
	\begin{tikzpicture}
	\node[cross out,draw=black] (x) at (0,0) {};
	\node[shape=circle,draw=black] (A) at (0,0) {};
	\node[shape=circle,draw=black] (B) at (.5,-.3) {};
	\node[shape=circle,draw=black,fill=blue!20] (C) at (.5,.3) {};
	\path [->] (A) edge (B);
	\path [->] (A) edge (C);
	\path [->] (B) edge (C);
	\end{tikzpicture}
	\end{array}}(t) & = (1 - e^{-t}) (1 - (1+ t)e^{-t}) \,, 
	\\
	{\scriptsize 
	\begin{array}{c}
	\begin{tikzpicture}
	\node[cross out,draw=black] (x) at (0,0) {};
	\node[shape=circle,draw=black] (A) at (0,0) {};
	\node[shape=circle,draw=black,fill=blue!20] (B) at (.5,-.4) {};
	\node[shape=circle,draw=black,fill=blue!20] (C) at (.5,.4) {};
	\node[shape=circle,draw=black,fill=blue!20] (D) at (.5,0) {};
	\path [->] (A) edge node[left] {} (B);
	\path [->] (A) edge node[left] {} (C);
	\path [->] (A) edge node[left] {} (D);
	\end{tikzpicture}
	\end{array}} (t) & = (1-e^{-t})^3 \,,  
	\\
    {\scriptsize 
	\begin{array}{c}
	\begin{tikzpicture}
	\node[cross out,draw=black] (x) at (0,0) {};
	\node[shape=circle,draw=black] (A) at (0,0) {};
	\node[shape=circle,draw=black] (B) at (.5,0) {};
	\node[shape=circle,draw=black,fill=blue!20] (C) at (1,0) {};
	\node[shape=circle,draw=black] (D) at (0.25,0.4) {};
	\path [->] (A) edge node[left] {} (B);
	\path [->] (B) edge node[left] {} (C);
	\path [<-] (B) edge node[left] {} (D);
	\path [<-] (D) edge (A);
	\end{tikzpicture}
	\end{array}}(t) = 
	{\scriptsize 
	\begin{array}{c}
	\begin{tikzpicture}
	\node[cross out,draw=black] (x) at (0,0) {};
	\node[shape=circle,draw=black] (A) at (0,0) {};
	\node[shape=circle,draw=black] (B) at (.5,0) {};
	\node[shape=circle,draw=black,fill=blue!20] (C) at (1,0) {};
	\node[shape=circle,draw=black] (D) at (.75,.4) {};
	\path [->] (A) edge node[left] {} (B);
	\path [->] (B) edge node[left] {} (C);
	\path [->] (B) edge node[left] {} (D);
	\path [->] (D) edge (C);
	\end{tikzpicture}
	\end{array}}(t) = - {\scriptsize
	\begin{array}{c}
	\begin{tikzpicture}
	\node[cross out,draw=black] (x) at (0,0) {};
	\node[shape=circle,draw=black] (A) at (0,0) {};
	\node[shape=circle,draw=black] (A1) at (0.5,0) {};
	\node[shape=circle,draw=black] (B) at (1,0.3) {};
	\node[shape=circle,draw=black,fill=blue!20] (C) at (1,-.3) {};
	\node[shape=circle,draw=black,fill=blue!20] (D) at (1.5,0.3) {};
	\path [->] (A) edge node[left] {} (A1);
	\path [->] (A1) edge node[left] {} (B);
	\path [->] (A1) edge node[left] {} (C);
	\path [->] (B) edge (D);
	\end{tikzpicture}
	\end{array}} (t) & = -1 + (-1 + 2t +\tfrac12 t^2 )e^{-t} + (2+t) e^{-2t} \,, 
	\\
	{\scriptsize 
	\begin{array}{c}
	\begin{tikzpicture}
	\node[cross out,draw=black] (x) at (0,0) {};
	\node[shape=circle,draw=black] (A) at (0,0) {};
	\node[shape=circle,draw=black] (B) at (.5,0) {};
	\node[shape=circle,draw=black,fill=blue!20] (C) at (1,-0.25) {};
	\node[shape=circle,draw=black,fill=blue!20] (D) at (1,0.25) {};
	\node[shape=circle,draw=black,fill=blue!20] (E) at (.25,0.4) {};
	\path [->] (A) edge node[left] {} (B);
	\path [->] (A) edge node[left] {} (E);
	\path [->] (B) edge node[left] {} (C);
	\path [->] (B) edge node[left] {} (D);
	\end{tikzpicture}
	\end{array}} (t) = - {\scriptsize 
	\begin{array}{c}
	\begin{tikzpicture}
	\node[cross out,draw=black] (x) at (0,0) {};
	\node[shape=circle,draw=black] (A) at (0,0) {};
	\node[shape=circle,draw=black] (B) at (.5,0) {};
	\node[shape=circle,draw=black,fill=blue!20] (C) at (1,0) {};
	\node[shape=circle,draw=black,fill=blue!20] (D) at (0.25,0.4) {};
	\path [->] (A) edge node[left] {} (B);
	\path [->] (B) edge node[left] {} (C);
	\path [->] (B) edge node[left] {} (D);
	\path [<-] (D) edge (A);
	\end{tikzpicture}
	\end{array}}(t)  & = (1-e^{-t}) (1 - 2t e^{-t} - e^{-2t}) \,, 
	\\
	{\scriptsize 
	\begin{array}{c}
	\begin{tikzpicture}
	\node[cross out,draw=black] (x) at (0,0) {};
	\node[shape=circle,draw=black] (A) at (0,0) {};
	\node[shape=circle,draw=black] (B) at (.5,0) {};
	\node[shape=circle,draw=black,fill=blue!20] (C) at (0.4,-0.4) {};
	\node[shape=circle,draw=black,fill=blue!20] (D) at (0.4,0.4) {};
	\node[shape=circle,draw=black,fill=blue!20] (E) at (1,0) {};
	\path [->] (A) edge node[left] {} (B);
	\path [->] (A) edge node[left] {} (C);
	\path [->] (B) edge node[left] {} (E);
	\path [<-] (D) edge (A);
	\end{tikzpicture}
	\end{array}}(t) = - {\scriptsize 
	\begin{array}{c}
	\begin{tikzpicture}
	\node[cross out,draw=black] (x) at (0,0) {};
	\node[shape=circle,draw=black] (A) at (0,0) {};
	\node[shape=circle,draw=black] (B) at (.8,0) {};
	\node[shape=circle,draw=black,fill=blue!20] (C) at (0.4,-0.4) {};
	\node[shape=circle,draw=black,fill=blue!20] (D) at (0.4,0.4) {};
	\path [->] (A) edge node[left] {} (B);
	\path [->] (A) edge node[left] {} (C);
	\path [->] (B) edge node[left] {} (D);
	\path [<-] (D) edge (A);
	\end{tikzpicture}
	\end{array}}(t) & =  (1 - e^{-t})^2 (1 - (1+ t)e^{-t}) 
	\\
	{\scriptsize
	\begin{array}{c}
	\begin{tikzpicture}
	\node[cross out,draw=black] (x) at (0,0) {};
	\node[shape=circle,draw=black] (A) at (0,0) {};
	\node[shape=circle,draw=black] (B) at (.5,0.3) {};
	\node[shape=circle,draw=black,fill=blue!20] (C) at (.5,-.3) {};
	\node[shape=circle,draw=black] (D) at (1,0.3) {};
	\node[shape=circle,draw=black,fill=blue!20] (E) at (1.5,0.3) {};
	\path [->] (A) edge node[left] {} (B);
	\path [->] (A) edge node[left] {} (C);
	\path [->] (B) edge (D);
	\path [->] (D) edge (E);
	\end{tikzpicture}
	\end{array}} (t) = 
	- {\scriptsize 
	\begin{array}{c}
	\begin{tikzpicture}
	\node[cross out,draw=black] (x) at (0,0) {};
	\node[shape=circle,draw=black] (A) at (0,0) {};
	\node[shape=circle,draw=black] (B) at (.8,0) {};
	\node[shape=circle,draw=black,fill=blue!20] (C) at (0.4,-0.4) {};
	\node[shape=circle,draw=black] (D) at (0.4,0.4) {};
	\path [->] (A) edge node[left] {} (C);
	\path [<-] (C) edge node[left] {} (B);
	\path [<-] (B) edge node[left] {} (D);
	\path [<-] (D) edge (A);
	\end{tikzpicture}
	\end{array}}(t) & =  (1 -e^{-t} ) (1 - (1+ t + \tfrac12 t^2)e^{-t} ) 
\\
		{\scriptsize
	\begin{array}{c}
	\begin{tikzpicture}
	\node[cross out,draw=black] (x) at (0,0) {};
	\node[shape=circle,draw=black] (A) at (0,0) {};
	\node[shape=circle,draw=black] (B) at (.5,0.3) {};
	\node[shape=circle,draw=black] (C) at (.5,-.3) {};
	\node[shape=circle,draw=black,fill=blue!20] (D) at (1,0.3) {};
	\node[shape=circle,draw=black,fill=blue!20] (E) at (1,-0.3) {};
	\path [->] (A) edge node[left] {} (B);
	\path [->] (A) edge node[left] {} (C);
	\path [->] (B) edge (D);
	\path [->] (C) edge (E);
	\end{tikzpicture}
	\end{array}} (t) =  - {\scriptsize 
	\begin{array}{c}
	\begin{tikzpicture}
	\node[cross out,draw=black] (x) at (0,0) {};
	\node[shape=circle,draw=black] (A) at (0,0) {};
	\node[shape=circle,draw=black,fill=blue!20] (B) at (.8,0) {};
	\node[shape=circle,draw=black] (C) at (0.4,-0.4) {};
	\node[shape=circle,draw=black] (D) at (0.4,0.4) {};
	\path [->] (A) edge node[left] {} (C);
	\path [->] (C) edge node[left] {} (B);
	\path [<-] (B) edge node[left] {} (D);
	\path [<-] (D) edge (A);
	\end{tikzpicture}
	\end{array}} (t) & =  (1 - (1+ t)e^{-t} )^2 
	\\
	{\scriptsize 
	\begin{array}{c}
	\begin{tikzpicture}
	\node[cross out,draw=black] (x) at (0,0) {};
	\node[shape=circle,draw=black] (A) at (0,0) {};
	\node[shape=circle,draw=black] (B) at (.5,0) {};
	\node[shape=circle,draw=black] (C) at (1,0) {};
	\node[shape=circle,draw=black] (D) at (1.5,0) {};
	\node[shape=circle,draw=black,fill=blue!20] (E) at (2,0) {};
	\path [->] (A) edge node[left] {} (B);
	\path [->] (B) edge node[left] {} (C);
	\path [->] (C) edge (D);
	\path [->] (D) edge (E);
	\end{tikzpicture}
	\end{array}}(t) & = 1 - (1+ t + \tfrac12 t^2 + \tfrac16 t^3 )e^{-t} \,, 
	\\
	{\scriptsize 
	\begin{array}{c}
	\begin{tikzpicture}
	\node[cross out,draw=black] (x) at (0,0) {};
	\node[shape=circle,draw=black] (A) at (0,0) {};
	\node[shape=circle,draw=black] (A1) at (0.5,0) {};
	\node[shape=circle,draw=black] (B) at (1,0) {};
	\node[shape=circle,draw=black,fill=blue!20] (C) at (1.5,-0.25) {};
	\node[shape=circle,draw=black,fill=blue!20] (D) at (1.5,0.25) {};
	\path [->] (A) edge node[left] {} (A1);
	\path [->] (A1) edge node[left] {} (B);
	\path [->] (B) edge node[left] {} (C);
	\path [->] (B) edge node[left] {} (D);
	\end{tikzpicture}
	\end{array}} (t) & =  1 - (2 + t^2 )e^{-t}  + e^{-2t} \,, 
	\\
	{\scriptsize 
	\begin{array}{c}
	\begin{tikzpicture}
	\node[cross out,draw=black] (x) at (0,0) {};
	\node[shape=circle,draw=black] (A0) at (0,0) {};
	\node[shape=circle,draw=black] (A) at (0.5,0) {};
	\node[shape=circle,draw=black,fill=blue!20] (B) at (1,-.4) {};
	\node[shape=circle,draw=black,fill=blue!20] (C) at (1,.4) {};
	\node[shape=circle,draw=black,fill=blue!20] (D) at (1,0) {};
	\path [->] (A0) edge (A);
	\path [->] (A) edge node[left] {} (B);
	\path [->] (A) edge node[left] {} (C);
	\path [->] (A) edge node[left] {} (D);
	\end{tikzpicture}
	\end{array}} (t) & = 1 + (\tfrac32 - 3t) e^{-t}   - 3 e^{-2t} + \tfrac12 e^{-3t} \,,
	\\
	{\scriptsize 
	\begin{array}{c}
	\begin{tikzpicture}
	\node[cross out,draw=black] (x) at (0,0) {};
	\node[shape=circle,draw=black] (A) at (0,0) {};
	\node[shape=circle,draw=black,fill=blue!20] (B) at (.5,-.5) {};
	\node[shape=circle,draw=black,fill=blue!20] (C) at (.5,.5) {};
	\node[shape=circle,draw=black,fill=blue!20] (D) at (.5,0.17) {};
	\node[shape=circle,draw=black,fill=blue!20] (E) at (.5,-0.17) {};
	\path [->] (A) edge node[left] {} (B);
	\path [->] (A) edge node[left] {} (C);
	\path [->] (A) edge node[left] {} (D);
	\path [->] (A) edge node[left] {} (E);
	\end{tikzpicture}
	\end{array}} (t) & = (1-e^{-t})^4
\end{align*}
\end{widetext}

\begin{acknowledgments}
WM wishes to thank Clelia de Mulatier for useful discussions and acknowledges the support from the NWO Klein grant awarded to NWA route 2 in 2020.
\end{acknowledgments}


%

\end{document}